\documentclass[%
reprint,
superscriptaddress,
 amsmath,amssymb,
 aps,
]{revtex4-2}
\usepackage{multirow}
\usepackage{graphicx}
\usepackage{dcolumn}

\usepackage{bm}

\usepackage{xcolor}

\usepackage{lipsum}

\usepackage{hyperref}
\hypersetup{
    colorlinks=true,
    linkcolor=cyan,
    filecolor=black,      
    urlcolor=black,
    citecolor=black,
    }

\makeatletter
\renewcommand*{\@fnsymbol}[1]{\ensuremath{\ifcase#1\or *\or \dagger\or \ddagger\or
   \mathsection\or \mathparagraph\or \|\or **\or \dagger\dagger
   \or \ddagger\ddagger \else\@ctrerr\fi}}
\makeatother

\begin{document}

\preprint{AIP/123-QED}

\title{Coherent control of the translational and point group symmetries of crystals with light}

\author{Guru Khalsa}
\email[]{guru.khalsa@cornell.edu}
\thanks{These authors contributed equally to this work.}
\affiliation{Department of Materials Science and Engineering, Cornell University, Ithaca, New York 14853, USA}%

\author{Jeffrey Z. Kaaret}
\thanks{These authors contributed equally to this work.}
\affiliation{School of Applied and Engineering Physics, Cornell University, Ithaca, New York 14853, USA}%

\author{Nicole A. Benedek}
\email[]{nbenedek@cornell.edu}
\affiliation{Department of Materials Science and Engineering, Cornell University, Ithaca, New York 14853, USA}%


\begin{abstract}
We use theory and first-principles calculations to explore mechanisms for control of the translational and point group symmetries of crystals in ultrafast optical experiments. We focus in particular on mechanisms that exploit anharmonic (biquadratic) lattice couplings between a driven infrared-active phonon mode and other modes at arbitrary wave vector, which are always allowed by symmetry in any space group. We use Floquet theory to develop a general phase diagram depicting the various dynamical regimes accessible to materials, with simulated dynamics to illustrate how the biquadratic coupling changes materials structure depending on both extrinsic factors (light pulse characteristics) and intrinsic materials parameters (phonon frequencies, phonon coupling strengths). We use our phase diagram, in conjunction with density functional theory calculations, both to suggest experiments to reveal hidden structural order in perovskite KTaO$_3$, and to provide additional insights into recently reported experiments on SrTiO$_3$ and LiNbO$_3$.
\end{abstract}

\pacs{Valid PACS appear here}
\maketitle

\section{Introduction} \label{sec:introduction}
Phase transitions in crystals are often characterized in terms of symmetry changes. For example, structural phase transitions, which involve some change to the symmetry of the lattice, are ubiquitous in some classes of materials, complex oxides in particular, and have been studied extensively for many decades \cite{Goldschmidt1926,Landau1965,Shirane1974,Li2021}. Phase transitions involving broken time-reversal symmetry give rise to magnetic materials and have been similarly well-studied \cite{Landau1965,toledano1987}. In recent years, attention has turned to more exotic phase transitions. Broken rotational symmetries of the electronic states (but not seen in the lattice) give rise to strong correlations in electronic nematic systems \cite{Borzi2007,Gupta2021}. The relative twist between layers of stacked two-dimensional materials, such as graphene, controls the overall point group and translational symmetry of the system and gives rise to significant changes in the density of states and corresponding electronic properties, depending on the twist angle \cite{Cao2018}. In each of these cases, detailed study of the relevant phase transitions has revealed new physical insights and advanced our fundamental understanding of condensed matter.

In addition to their fundamental importance, materials that undergo phase transitions are of great practical interest because they provide opportunities for the control of functional properties with external fields. Control of the polarization with electric fields in ferroelectric materials is exploited in certain types of random access memory \cite{Zhang_2012}. The key component of many sensors and actuators is a piezoelectric material \cite{Jiao2020}, which exhibits large changes in its electrical polarization in response to external stress \cite{piezo1880} (and \textit{vice versa} \cite{Lippmann_1881}). The challenge for condensed  matter and materials scientists is that the property of interest may not always couple directly with an external field. For example, the magnetic and electronic properties of many ABO$_3$ perovskites oxides (where A and B are cations and O is oxygen) are associated with lattice (phonon) modes that do not couple directly to external fields. It is possible to identify mechanisms that couple these modes by symmetry to others that are, say, electric field-controllable thereby giving indirect control of properties with an external field \cite{bousquet08,MAH8,benedek11,benedek12,benedek22}. However, not all materials exhibit the required crystallographic symmetries for such mechanisms.

The development of bright mid-infrared and THz laser sources, capable of resonantly exciting phonons in crystals, has created new opportunities for the control of functional properties with external fields, namely light. One such mechanism involves resonantly exciting an infrared (IR)-active phonon mode of the crystal to large amplitude (Q$_{IR}$), which induces quasi-static displacements of some Raman-active modes (Q$_{R}$) \textit{via} anharmonic coupling of the form Q$_{IR}^2$Q$_R$. In these so-called \emph{nonlinear phononics} experiments, the light pulse can induce a transition to a metastable phase with properties that are either difficult or impossible to access in the equilibrium structure at a given temperature, for example, metal-insulator phase transitions \cite{rini07},  superconductivity \cite{Fausti2011, mankowsky14}, and changes in orbital ordering \cite{Miller2015}. In most of the experiments that have been reported so far, the optically excited IR mode couples to Raman modes that are also at the Brillouin zone center, such that the induced metastable phase has the same translational symmetry as the ground-state structure. Is it possible to identify other anharmonic lattice couplings, involving phonon modes at non-zero wave vector, which would allow us to dynamically stabilize phases with a translational symmetry different to that of the parent ground-state structure?

\onecolumngrid
\clearpage
\begin{widetext}
    \begin{figure}[h!]
    \centering
    \includegraphics[width=.88\textwidth]{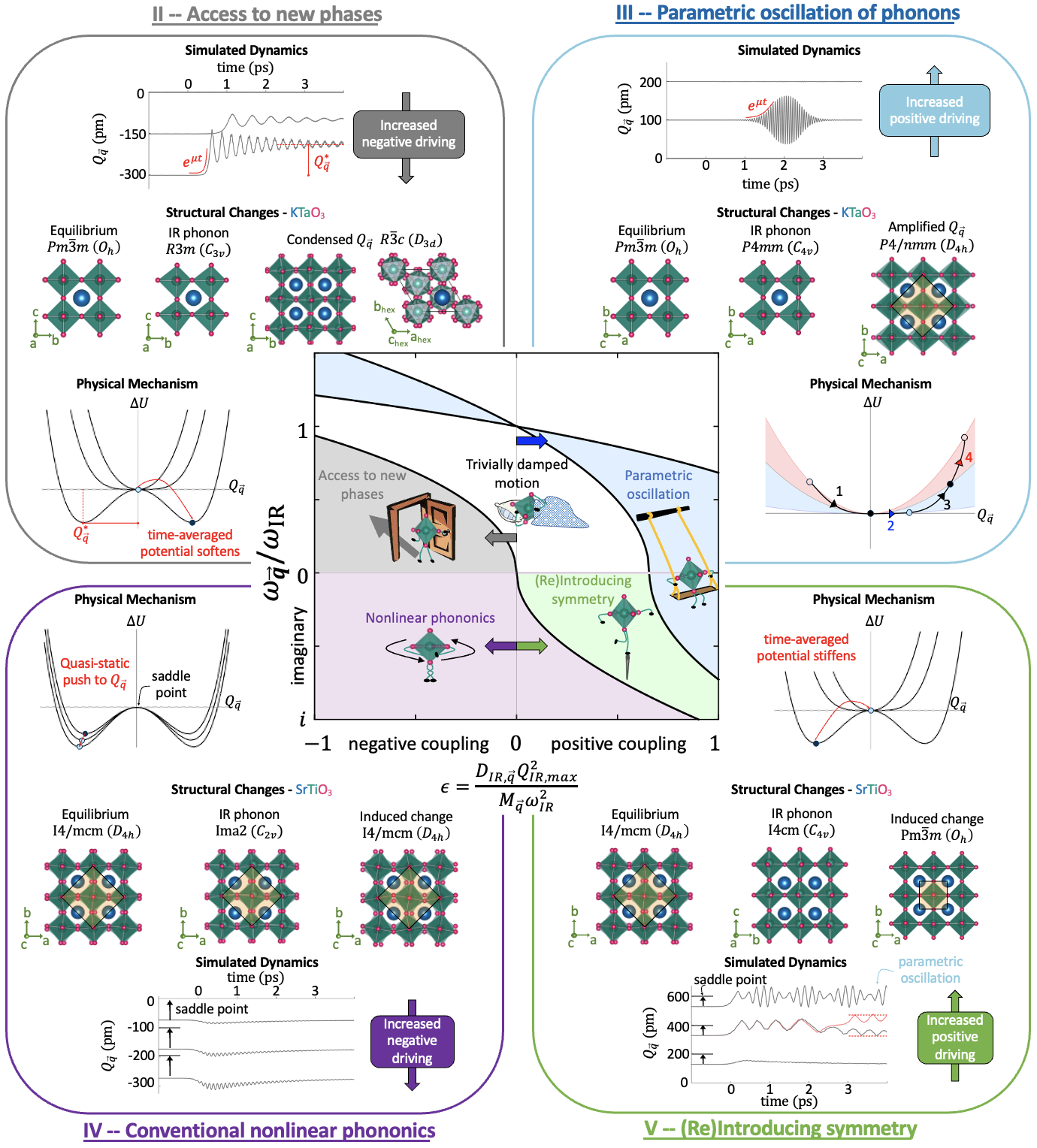}
    \caption{Dynamical regimes accessible through biquadratic coupling between an optically pumped IR-active phonon, $Q_{IR}$, and a mode at arbitrary wave vector, $Q_{\vec{q}}$. The Floquet phase diagram (center) is shown as a function of the frequency ratio of the biquadratically coupled mode ($\omega_{\vec{q}}$) to the excited IR mode ($\omega_{IR}$) versus the driving strength $\epsilon \propto D_{IR,\vec{q}} Q_{IR}^2$ (see full definition below Eqn. \ref{eqn:hill_matrix}). The simulated dynamics, structural changes, and schematic mechanisms are shown in the color-coordinated panels. When $Q_{\vec{q}}$ has a positive force constant (real frequency) at equilibrium (top-half, $\omega_{\vec{q}}/\omega_{IR} > 0$), trivially damped motion is expected in much of the phase diagram (I -- white). For negative biquadratic coupling ($D_{IR,\vec{q}} <0$), $Q_{\vec{q}}$ can be frozen in by optical excitation of an IR phonon, leading to novel transient structural phases (II -- gray). Near $\omega_{\vec{q}}/\omega_{IR} = 1$, parametric oscillation of $Q_{\vec{q}}$ is possible (III -- blue). In the purple region of the phase diagram (Region IV) we imagine a high-symmetry reference structure for the material of interest, which may be virtual, located at the saddle point (see inset). $Q_{\vec{q}}$ has a negative force constant in this phase and drives a non-transient structural transition to a lower-symmetry phase. In the low-symmetry phase $Q_{\vec{q}}$ can be displaced quasi-statically by optical excitation of $Q_{IR}$; this is the conventional nonlinear phononics effect. For positive coupling between $Q_{\vec{q}}$ and the IR mode ($D_{IR,\vec{q}} >0$), the symmetry broken by $Q_{\vec{q}}$ can be (re)introduced into the structure (V -- green), beyond which $Q_{\vec{q}}$ again parametrically oscillates.}
    \label{fig:phase_diagram_expanded}
\end{figure}
\end{widetext}
\twocolumngrid


In this work, we use theory and first-principles calculations to explore and elucidate anharmonic lattice coupling pathways involving a zone-center IR-active mode and modes at arbitrary wave vector, Q$_{\vec{q}}$. We show that biquadratic coupling between these modes, Q$^2_{IR}$Q$^2_{\vec{q}}$, which is \emph{always} allowed by symmetry in \emph{any} space group, can be exploited in ultrafast optical experiments such that resonant excitation of an IR-active mode at the zone center can dynamically induce phases with translational symmetry different to that of the ground state through coupling to and condensation of modes at arbitrary (non-zero) wave vector. This biquadratic coupling potentially offers a new pathway through which the functional properties of materials can be dynamically controlled. We use classical Floquet theory to map out a general phase diagram showing the various dynamical regimes that may be accessed with this biquadratic coupling in modern ultrafast optical experiments; this phase diagram is our key result and is shown in Fig. \ref{fig:phase_diagram_expanded}. Our approach highlights how the nonlinear phononics mechanism can be viewed as a specific example of a more general parametric amplification process and reveals dynamical regimes that may be missed in the time-averaged theories commonly used to study these processes. We show how different parts of the phase diagram can be accessed by tuning the light polarization, frequency and peak electric field (extrinsic experimental parameters), and the relative frequencies and strength of the coupling between the pumped IR and coupled modes (determined by intrinsic materials parameters, such as the crystal structure and chemical composition). The phase diagram thus functions as a tool that can be used to interpret existing experiments (such as recent work demonstrating transient -- that is, not long-lasting or deterministic -- polarization switching in LiNbO$_3$ \cite{mankowsky17}) and to design future experiments. We use our phase diagram, in combination with first-principles DFT calculations, to both interpret the results of a recently reported experiment on SrTiO$_3$ showing light control of translational symmetry \cite{fechner2023}, and to suggest future experiments that can access hidden structural order in perovskite KTaO$_3$. Our work goes further than previously published studies by exploring a much greater range of dynamical regimes within a single coherent framework.

\section{Theoretical Model}
\subsection{Simple model of biquadratic coupling} \label{sec:simple_model}
We start by writing down a general equation for describing the response of a centrosymmetric crystal to resonant excitation of an IR-active phonon mode by a short, intense mid-IR pulse. We consider the case where the dominant dynamics induced are those of the excited IR mode, $Q_{IR}$, and another mode coupled to it at arbitrary wave vector $\vec{q}$, $Q_{\vec{q}}$. The lattice potential energy is assumed to be:

\begin{widetext}

\begin{equation} \label{eqn:Potential_energy}
    \begin{split}
        U  = \frac{1}{2}K_{IR}Q_{IR}^2 + \frac{1}{2}K_{\vec{q}}Q_{\vec{q}}^2 +D_{IR,\vec{q}}Q_{IR}^2Q_{\vec{q}}^2 +\frac{1}{4} D_{\vec{q}} Q_{\vec{q}}^4-\Delta \vec{P}\cdot \vec{E} ,
    \end{split}
\end{equation}
\end{widetext}

\noindent
where $K_{IR}$ and $K_{\vec{q}}$ are the force constants of the relevant phonon modes at harmonic order,  $D_{\vec{q}}$ is the fourth-order force constant for $Q_{\vec{q}}$, and $D_{IR,\vec{q}}$ is the biquadratic coupling coefficient coupling $Q_{IR}$ and $Q_{\vec{q}}$. The polarization change in the crystal due to the electric field $\vec{E}$ of the light pulse is given to linear order by $\Delta\vec{P} = \tilde{Z}^* Q_{IR}$, where $\tilde{Z}^*$ is the mode-effective charge of the excited IR mode (as defined in \cite{gonze1997dynamical}). In previous work, we showed that higher-order terms in the polarization are important for understanding changes in the optical properties of materials due to optical excitation \cite{Khalsa2021}; we ignore those terms in this work to simplify the analysis, and because here we are focused on understanding \emph{structural} changes due to optical excitation. Also note that we ignore explicit electron-phonon interactions -- the electrons adiabatically follow the excited phonon coordinates and the equations of motion are classical. Recent work \cite{Caruso2023} on the quantum theory of lattice dynamics in the presence of external driving fields has validated this classical approach; the framework developed here could also be adapted to incorporate the quantum theory of Ref. \onlinecite{Caruso2023}.

As mentioned above, much of the previous work on nonlinear phononics has focused on systems where the excited IR mode is coupled to Raman-active modes at the zone center through a term of the form $Q_{IR}^2 Q_{R}$. Whether or not such a term is an allowed invariant in the lattice potential is determined by crystallographic symmetry and hence only some materials are candidates for nonlinear phononics experiments that exploit this type of cubic anharmonic coupling. In addition, since the term $Q_{IR}^2 Q_{R}$ involves modes at the zone center only, 
there is no change in the translational symmetry of the crystal upon optical excitation. In contrast, the biquadratic term $Q_{IR}^2 Q_{\vec{q}}^2$ shown in Eqn. \ref{eqn:Potential_energy} is \emph{always} allowed by symmetry between \emph{all} modes and in \emph{all} space groups (biquadratic coupling between $Q_{IR}$ and a phonon at $\vec{q}=\vec{0}$ has been explored in previous work \cite{Subedi2014}).  How can this biquadratic coupling be exploited for light control of properties? By rearranging terms in Eqn. \ref{eqn:Potential_energy}, we see that the effective force constant of $Q_{\vec{q}}$ ($\tilde{K}_{\vec{q}}$) is renormalized by the motion of the IR mode:

\begin{equation} \label{eqn:ZE_Force_constant}
    \begin{split}
        \tilde{K}_{\vec{q}}  &= K_{\vec{q}}  +2D_{IR,\vec{q}}Q_{IR}^2.
    \end{split}
\end{equation}

\noindent
Displacing the IR mode with ${D_{IR,\vec{q}}>0}$ increases the force constant $\tilde{K}_{\vec{q}}$, thereby stiffening this mode. In contrast, when $D_{IR,\vec{q}}<0$ the effective force constant $\tilde{K}_{\vec{q}}$ decreases. If the IR mode displacement and the magnitude of the biquadratic coupling are sufficiently large compared to the equilibrium force constant $K_{\vec{q}}$ (that is, when $K_{\vec{q}} \leq 2D_{IR,\vec{q}}Q_{IR}^2$), then $Q_{\vec{q}}$ will `freeze in', resulting in a new structural phase with symmetry broken by $Q_{IR}$ and $Q_{\vec{q}}$. Exciting IR modes to large amplitude is now possible with current bright mid-IR and THz sources, but the IR motion is oscillatory. Can oscillatory motion of the IR mode induce new \emph{transient} structural phases via biquadratic coupling to $Q_{\vec{q}}$? In what follows, we use Floquet theory, first-principles calculations, and dynamical simulations to uncover new dynamical regimes accessible through the biquadratic coupling in Eqn. \ref{eqn:Potential_energy}.

\onecolumngrid
\subsection{Floquet Theory}\label{sec:Floquet}

Consider the equations of motion that are derived from Eqn. \ref{eqn:Potential_energy} by taking derivatives with respect to the mode coordinates, $Q_{IR}$ and $Q_{\vec{q}}$. We find,

\twocolumngrid
\begin{widetext}
\begin{equation}\label{eqn:EOM}
    \begin{split}
          \ddot{Q}_{IR} + 2 \gamma_{IR} \dot{Q}_{IR} + \frac{1}{M_{IR}}\left(K_{IR} + 2 D_{IR,\vec{q}} Q_{\vec{q}}^2\right) Q_{IR}         = \frac{\tilde{Z}^*}{M_{IR} } E,  \\ 
         \ddot{Q}_{\vec{q}} + 2 \gamma_{\vec{q}} \dot{Q}_{\vec{q}} + \frac{1}{M_{\vec{q}}}\left(K_{\vec{q}} + 2 D_{IR,\vec{q}} Q_{IR}^2\right) Q_{\vec{q}} + \frac{D_{\vec{q}}}{M_{\vec{q}} }Q_{\vec{q}}^3
         = 0.
    \end{split}
\end{equation}
\end{widetext}

\noindent To simplify the analysis we ignore all terms that are higher than harmonic order except for the biquadratic coupling and reintroduce the $D_{\vec{q}}Q_{\vec{q}}^4$ term only when necessary to guarantee a finite minimum in $U$ with respect to $Q_{\vec{q}}$. Additionally, we assume that the pumped IR mode is undamped ($\gamma_{IR} = 0$), thereby allowing for the exploration of the long timescale dynamics. We justify this assumption \emph{a posteriori} by numerical exploration of the dynamics, finding that our simulated short-timescale dynamics of $Q_{\vec{q}}$ are comparable to the Floquet results (which assume continuous periodic driving). 

After resonant excitation of the IR mode with a Gaussian pulse we find sinusoidal periodic motion for $Q_{IR}$ of the form

\begin{equation}\label{eqn:IR_approx} 
    Q_{IR}\left( t \right) = 2 \frac{\tilde{Z}^*\tilde{E}}{K_{IR}} cos\left({\omega_{IR} t  + \phi}\right).
\end{equation}

\noindent
Here $\tilde{E}$ is defined as $\eta E_{0} \tau f_{IR}$, where $E_0$ is the peak electric field, $\tau$ is the full-width at half-maximum of the electric field, $f_{IR}$ is the (linear) IR frequency, and $\eta$ characterizes the shape of the pulse (for a Gaussian pulse ${\eta = \sqrt{\frac{(\pi/2)^3}{2 \ln(2)} } \approx 1.67}$). Then $ \eta \tau f_{IR}$ measures the number of cycles the IR mode is driven by the electric field of the light pulse and the peak IR displacement is $Q_{IR,0} = 2\frac{\tilde{Z}^*\tilde{E}}{K_{IR}}$. We neglect $\phi$ in what follows. $Q_{IR}$ is therefore a periodic driver for the general lattice motion. This can be contrasted with other recent Floquet approaches, where off-resonant periodic excitation of the \emph{electronic} states is motivating the Floquet approach and enabling access to novel nonequilibrium phases \cite{claassen2017,Oka2019,delatorre21}.

Eqn. \ref{eqn:IR_approx} assumes that the biquadratic coupling between $Q_{IR}$ and $Q_{\vec{q}}$ is zero ($D_{IR,\vec{q}} = 0$). This assumption can be justified from a perturbation theory perspective, since $Q_{\vec{q}}$ is initially characterized by small fluctuations about zero amplitude. However, it obviously will not hold when $Q_{\vec{q}}$ is displaced to large amplitudes. In this scenario, the force constant of the IR mode can also be renormalized by $Q_{\vec{q}}$, $\Delta K_{IR} \approx 2 D_{IR,\vec{q}} Q_{\vec{q}}^2 $, such that the IR mode also freezes in. Using Eqn. \ref{eqn:IR_approx} for $Q_{IR}$ in the second line of Eqn. \ref{eqn:EOM}, rescaling time to $\theta = \omega_{IR}t$, and collecting terms, we find the following form,

\begin{equation}\label{eqn:hill}
    \begin{split}
        \frac{d}{d\theta} x(\theta) &= A(\theta) x(\theta),
    \end{split}
\end{equation}

\noindent
where we have defined the vector $x(\theta) = \left(Q_{\vec{q}},\frac{d}{d\theta}Q_{\vec{q}}\right)$ and the matrix 

\begin{equation} \label{eqn:hill_matrix}
    A(\theta) = \left( \begin{array}{cc}
    0 & 1 \\
    -\left(\delta + \epsilon + \epsilon cos \left(2\theta \right) \right) & -\nu \\
    \end{array} \right) .
\end{equation}
\noindent The dimensionless parameters ${\delta = \frac{1}{\omega_{IR}^2}\frac{K_{\vec{q}}}{M_{\vec{q}}} = \frac{\omega_{\vec{q}}^2}{\omega_{IR}^2}}$, ${\epsilon = D_{IR,\vec{q}} Q_{IR,0}^2/M_{\vec{q}} \omega_{IR}^2}$ (see Appendix \ref{app:units} for unit conversions) , and ${\nu = \frac{2\gamma_{\vec{q}}}{\omega_{IR}}}$ measure the square-frequency ratio, the driving strength, and effective damping for the driven mode $Q_{\vec{q}}$ due to the motion of $Q_{IR}$, respectively. The matrix $A(\theta)$ is periodic, with period $\pi$ (that is, $A(\theta + \pi) = A(\theta)$) and therefore represents a Hill's equation. The solutions to this equation (which may not be analytic), as well as how they depend on the parameters $\delta$, $\epsilon$ and $\nu$, can be obtained using the standard techniques of Floquet theory \cite{magnus2004hill,Rand2018}. 

Our primary interest is to find solutions to Eqn. \ref{eqn:hill} that predict exponential growth in $Q_{\vec{q}}$. These represent dynamical regimes in which optical pumping of an IR phonon induces large-amplitude changes in $Q_{\vec{q}}$, thereby driving a transient structural phase transition that could be resolved experimentally. We use the main results of our Floquet analysis to create a phase diagram, shown in the central panel of Fig. \ref{fig:phase_diagram_expanded}, from Eqn. \ref{eqn:hill} depicting these dynamical regimes (see Appendix \ref{app:Floquet_machinery} for more details).

\subsection{First-Principles Calculations and Dynamical Simulation} \label{sec:methods}

To connect the Floquet theory to real materials and experiments we find parameters for Eqns. \ref{eqn:EOM} \& \ref{eqn:hill_matrix} using density functional theory (DFT). Calculations were performed using VASP 6.2.0 \cite{VASP1,VASP2,VASP3}, using the projector augmented-wave (PAW) method in the local-density approximation (LDA) \cite{VASP_PAW}. The following states were included in the valence of the relevant PAW potentials: 3s$^2$3p$^6$4s$^1$ for K, 5p$^6$5d$^4$6s$^1$ for Ta, 4s$^2$4p$^6$5s$^2$ for Sr, 3s$^2$3p$^6$3d$^3$4s$^1$ for Ti, and 2s$^2$2p$^4$ for O. 
A force convergence tolerance of ${10^{-3}}$ eV/\AA\ was used for all calculations with a 4$\times$4$\times$4 Monkhorst-Pack \textbf{k}-point grid (for a 2$\times$2$\times$2 formula unit supercell of the primitive cubic perovskite unit cell) and plane-wave energy cutoffs of 600 eV (KTaO$_3$) and 700 eV (SrTiO$_3$). These values were chosen to converge  phonon frequencies, calculated with density functional perturbation theory \cite{Baroni2001DFPT} (DFPT), to within 5 cm$^{-1}$ when compared to incrementing the k-point grid to 8$\times$8$\times$8, and the energy cutoff to 800 eV for both KTaO$_3$ and SrTiO$_3$.
Our converged cubic lattice constants for KTaO$_3$ (3.959 \AA) and SrTiO$_3$ (3.859 \AA) are underestimated compared to the experimental lattice constants \cite{Vousden1951,ISHIHARA2004,Schmidbauer2012} of 3.988 \AA\ and 3.905 \AA, respectively, as expected in LDA.

We focus on pairs of IR modes and modes characterized by arbitrary wave vector that exhibit strong coupling at biquadratic order. To find these strongly coupled mode pairs, we calculated a series of phonon dispersion curves for structures in which an IR-active mode has been `frozen in' at varying amplitude. Modes that exhibit large changes in their frequencies as a function of IR mode amplitude were selected for further study. Quadratic, biquadratic, and quartic coupling coefficients were calculated via a series of symmetry-constrained frozen-phonon calculations. Mode-effective charges were calculated as defined in Ref. \cite{gonze1997dynamical}, which were found to be consistent with values obtained from modern theory of polarization calculations \cite{King-Smith1993, Resta1993, Vanderbilt1993}
conducted on +/-10 pm meshes of the IR-active phonons.
Non-equilibrium phonon dispersion curves were calculated with Quantum Espresso (plane-wave energy cutoff of 70 Ry, 8$\times$8$\times$8 Monkhorst-Pack \textbf{k}-point grid in the 5-atom cubic unit cell of KTaO$_3$), using Garrity-Bennett-Rabe-Vanderbilt pseudopotentials \cite{QuantumEspresso2009, GARRITY2014}, which confirmed the qualitative features of the induced dynamical instabilities, i.e. which modes develop imaginary frequencies when the IR mode amplitude is increased along a fixed polarization direction beyond a critical amplitude. 
These dispersion curves (Fig. \ref{fig:KTaO3_phonon_dispersion}) were calculated on a regular 10$\times$10$\times$10 $\mathbf{q}$-point grid, with the inclusion of a simple acoustic sum rule. Symmetry assignment of structural phases and irreducible representations of phonon modes were generated with the ISOTROPY Software Suite \cite{ISOTROPYGeneral, ISODISTORT}.

To explore the transient dynamics of the regions found in the Floquet analysis with numerical simulation of Eqns. \ref{eqn:EOM} \& \ref{eqn:hill_matrix}, we use a Runge-Kutta 5(4) method for numerical integration, as implemented in NumPy \cite{numpy}. Simulations were performed with Gaussian electric field pulses with duration $\tau=$ 500 fs, varied peak electric fields ($E_0$), IR phonon frequency, and peak field time set at $t=0$. First-principles calculations of the damping parameters $\gamma_{IR/\vec{q}}$ are computationally expensive. As a result, we have explored the dynamics as a function of damping parameter, ranging from 0 THz to 1 THz, showing only selected dynamics to highlight the response in different regions of the phase diagram. For initial conditions, $Q_{IR}$ is taken to be at rest long before the Gaussian pulse is present, and $Q_{\vec{q}}$ is given an amplitude of 0.1 pm to simulate weak fluctuation about the average equilibrium value. Note that all parameters are included when simulating Eqn. \ref{eqn:EOM}, in contrast to the numerical solution of Eqn. \ref{eqn:hill_matrix} where approximations have been made (see discussion in Sec. \ref{sec:Floquet}).

\section{Results and Discussion}
\subsection{Floquet phase diagram and dynamics}\label{sec:phase_diagram}

The phase diagram (center-panel of Fig. \ref{fig:phase_diagram_expanded}) shows different dynamical regimes for $Q_{\vec{q}}$. The vertical axis is the ratio of frequencies $\sqrt{\delta}=\omega_{\vec{q}}/ \omega_{IR}$, while the horizontal axis is ${\epsilon}$, the driving strength parameter $\left({D_{IR,\vec{q}} Q_{IR,0}^2/M_{\vec{q}} \omega_{IR}^2} \right)$. The sign of the anharmonic coupling $D_{IR,\vec{q}}$ is an intrinsic material property that dictates which side of the phase diagram is accessible -- positive coupling increases $\tilde{K}_{\vec{q}}$ (right) and negative coupling decreases $\tilde{K}_{\vec{q}}$ (left).

While the Floquet analysis informs us of regions of exponential growth or decay of $Q_{\vec{q}}$, it does not provide compact analytic results of the phase boundaries or dynamics in a given region that might inform future ultrafast experiments. We anticipate this need and include approximate analytic results, expressed in terms of microscopic materials parameters, in what follows and in Appendix \ref{app:Parametric}.

In the subsections below we discuss the basic physical mechanism in each region, followed by a materials example, and finally show the simulated dynamics of $Q_{\vec{q}}$. This follows the layout of the color-coded panels extending from the phase diagram in Fig. \ref{fig:phase_diagram_expanded}. Where appropriate we discuss previous theoretical and experimental work. We start with the top half of the phase diagram where the high-symmetry equilibrium structure is stable ($\sqrt{\delta} > 0$). 

\subsubsection{Region I -- Trivially damped motion}\label{sec:region_I}

For most of the upper-half of the phase diagram, trivial exponential decay is expected, suggesting that optically excited IR modes will not induce large amplitude dynamical behavior in $Q_{\vec{q}}$. Due to the exponentially decaying response of $Q_{IR}$ and $Q_{\vec{q}}$, a long-lived change in crystal symmetry is not expected. Recall that the biquadratic coupling pathway is allowed for all modes at all wave vectors. For weak driving of an IR phonon, our expectation is that most, or all of the coupled modes will be in this region. As can be seen from Fig. \ref{fig:phase_diagram_expanded}, a combination of external and intrinsic materials parameters ($\epsilon$) are needed to move one, or several, modes into another region of the phase diagram. We now focus on negative biquadratic coupling ($\epsilon<0$) and small $\sqrt{\delta}$ where, as mentioned in Sec. \ref{sec:simple_model}, a strong enough drive can decrease $\tilde{K}_{\vec{q}}$ such that $Q_{\vec{q}}$ freezes into the structure with a nonzero amplitude (Region II).

\subsubsection{Region II -- Access to new phases: Novel exponential growth and symmetry control}\label{sec:region_II}

Region II describes a region in which we expect $Q_{\vec{q}}$ to grow exponentially, thereby inducing a transient structural phase transition. Recent experimental work \cite{fechner2023} has shown that optical pumping of an IR phonon in the cubic phase of the perovskite SrTiO$_3$ at 135 K can induce a phonon mode that involves rotations of the TiO$_6$ octahedra, doubling the size of the unit cell and changing the translational symmetry. SrTiO$_3$ undergoes this structural phase transition with temperature at about 110 K -- optical excitation of an IR phonon effectively increases the transition temperature. As far as we are aware, this is the first experimental realization of this dynamical region. We show that in addition to changing the transition temperatures for structural phase transitions, it is also possible to induce structural phases that are not present \emph{at all} in the equilibrium phase diagram with temperature.

Taking $Q_{\vec{q}}(\theta) \propto e^{\mu\theta}$, with $\mu$ as a dimensionless exponential growth parameter ($\mu \omega_{IR}$ is the dimensional growth parameter), inserting this functional form in Eqn. \ref{eqn:hill}, and neglecting the $\cos(2\theta)$ term (which is equivalent to time averaging), we find the following relation

\begin{equation} \label{eqn:region_II_approx_growth}
    \mu^2 + \nu \mu +\left(\delta + \epsilon\right) = 0,
\end{equation}

\noindent which has the exponential growth solution ${\mu_{II,+} = \frac{1}{2}\left(-\nu + \sqrt{\nu^2 -4\left(\delta - |\epsilon| \right)} \right)}$ in Region II. This shows that exponential growth of $Q_{\vec{q}}$ in Region II is affected by damping, as well as the pulse characteristics through $\epsilon$. The phase boundary between Region I and Region II is defined by $\mu = 0$, which is solved when $|\epsilon| = \delta$. This result is surprisingly independent of the damping $\nu$, suggesting that although damping influences the growth rate of $Q_{\vec{q}}$ once in Region II it does not influence the location of the phase boundary. Comparing this analytic result with that from the Floquet analysis, we find that $|\epsilon| = \delta$ is a good approximation for the phase boundary. The analytic result works well for small $\delta$, \textit{i.e.} when $\omega_{\vec{q}} << \omega_{IR}$, but may slightly underestimate the value of $\epsilon$ needed to reach this regime (Supplementary Fig. \ref{fig:PhaseBoundary_approx}). This is attributed to neglecting the oscillatory component of the IR motion in the development of Eqn. \ref{eqn:region_II_approx_growth}.

To find the growth rate of $Q_{\vec{q}}$ in terms of microscopic parameters we need to rescale $\mu_{II,+}$ by $\omega_{IR}$  to account for the definition of dimensionless time $\theta = \omega_{IR} t$. The growth rate then becomes,

\begin{equation}
    \begin{split} \label{eqn:region_II_growth}
      \mu_{II,+} \omega_{IR} &= -\gamma_{\vec{q}} + \sqrt{  \gamma_{\vec{q}}^2 -\left( \omega_{\vec{q}}^2 - \frac{|D_{IR,\vec{q}}|}{M_{\vec{q}}} Q_{IR,0}^2  \right)   } 
    \end{split}.
\end{equation}

\noindent
Notice that on the phase boundary the terms in the parentheses of Eqn. \ref{eqn:region_II_growth} cancel, as a result $\mu_{II,+} = 0$. We can express the phase boundary in terms of microscopic parameters and pulse characteristics to derive a threshold field, which takes the form

\begin{equation}
    \begin{split} \label{eqn:crit_field}
      E_{0} \tau  &\geq \frac{1}{2} \frac{K_{IR}}{ \eta f_{IR}  \tilde{Z}^*} \sqrt{ \left|  \frac{K_{\vec{q}}}  { D_{IR,\vec{q}}  }\right| } .
    \end{split}
\end{equation}
\noindent
In this analysis, if $E_0\tau$ satisfies this inequality, $Q_{\vec{q}}$ grows exponentially, inducing a transient structural phase that changes the point group and translational symmetry of the crystal.

Beyond the threshold electric field, we can approximate the induced amplitude of  $Q_{\vec{q}}$ and its renormalized frequency by time-averaging Eqn. \ref{eqn:Potential_energy}. From this point of view, the curvature of the average potential energy landscape is decreasing and becoming negative with respect to $Q_{\vec{q}}$. That is, a minimum in the potential energy landscape develops beyond the threshold electric field in which $Q_{\vec{q}} \neq 0$. This new minimum survives the oscillatory $2\omega_{IR}$ component of the IR motion (Fig. \ref{fig:phase_diagram_expanded}, Region II). The motion of the IR phonon renormalizes both the amplitude of $Q_{\vec{q}}$,

\begin{equation}\label{eqn:new_minima}
    \begin{split}
           Q_{\vec{q},\pm}^*\left( Q_{IR} \right) &= \pm \sqrt{ | K_{\vec{q}}  +2D_{IR,\vec{q}} \left< Q_{IR}^2 \right> | /D_{\vec{q}}},
    \end{split}
\end{equation}

\noindent
and its frequency,

\begin{equation}\label{eqn:new_freq}
    \begin{split}
        \omega^*\left( Q_{IR} \right) &= \sqrt{2|\left(K_{\vec{q}}  +2D_{IR,\vec{q}} \left<Q_{IR}^2 \right>\right)|/M_{\vec{q}}},
    \end{split}
\end{equation}

\noindent
where we have explicitly shown the dependence of the new minimum and frequency on the IR phonon amplitude. 

The condensation of $Q_{\vec{q}}$ to nonzero amplitude modulates the existing crystal structure with a lengthscale derived from the wave vector $\vec{q}$. If $\vec{q} = \zeta_1^{-1} \vec{b}_1+\zeta_2^{-1}\vec{b}_2+\zeta_3^{-1}\vec{b}_3$ where $\vec{b}_i$ define the reciprocal lattice vectors of a lattice defined by $\vec{a}_i$ so that $\vec{a}_i \cdot \vec{b}_j = 2\pi\delta_{ij}$, the condensation of $Q_{\vec{q}}$ transiently induces a new periodicity to the crystal with modulation vector $\vec{\lambda} = \zeta_1 \vec{a}_1 +  \zeta_2 \vec{a}_2 +  \zeta_3 \vec{a}_3$, where each $\zeta_i$ may lead to (in)commensurability with its equilibrium lattice vector $\vec{a}_i$.

\begin{figure}[h]
    \includegraphics[width=.85\textwidth]{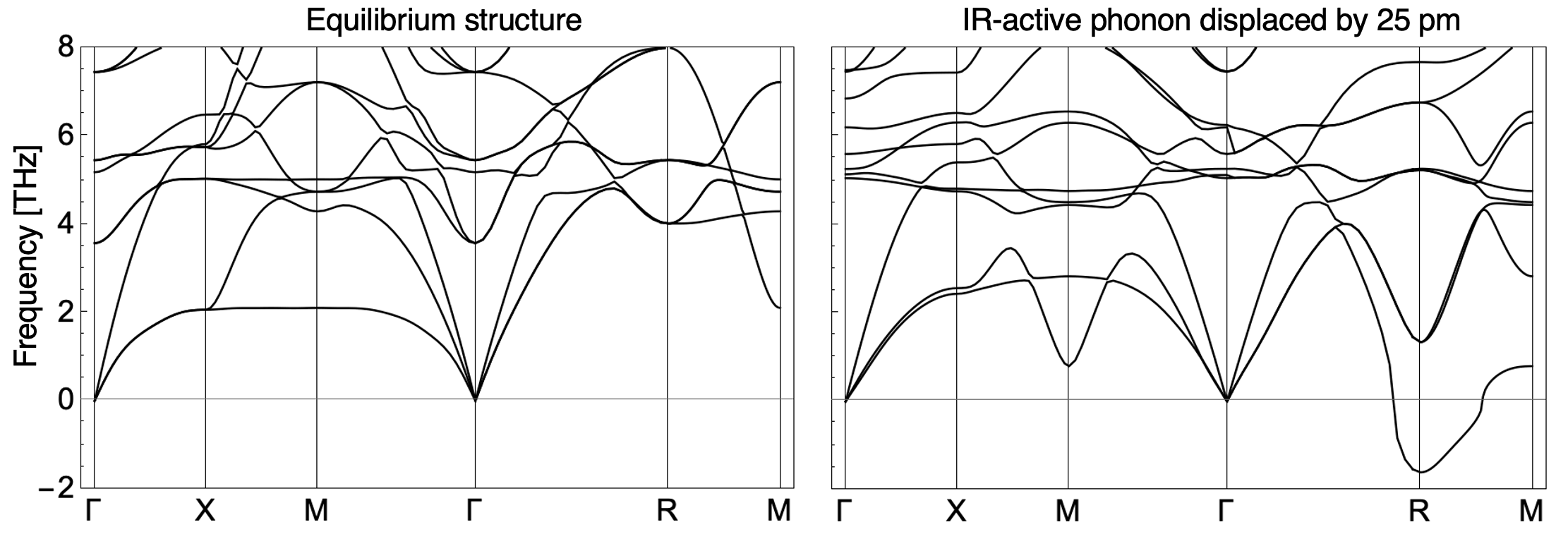}
    \caption{Phonon dispersion curves from our DFT calculations for KTaO$_3$ in the cubic perovskite structure at equilibrium (left) and with the 16.7 THz IR phonon displaced along [111] with an amplitude of 25 pm in the 5 atom cell (right). The IR phonon displacement splits and softens a branch at the R-point, leading to a driven instability of this mode. Modes with imaginary frequencies are depicted as having negative frequencies.}
    \label{fig:KTaO3_phonon_dispersion}
\end{figure}

We now turn to the example of KTaO$_3$, one of the few perovskites that remains cubic ($Pm\bar{3}m$ space group) at all temperatures; the 0 K phonon dispersion curve in Fig. \ref{fig:KTaO3_phonon_dispersion} shows all modes with real frequencies. KTaO$_3$ features three sets of triply degenerate IR-active optical phonons and no Raman-active phonons. As a result, new structural phases cannot be dynamically induced by relying on the conventional $Q_{IR}^2Q_R$ coupling. In contrast, every phonon at every wave vector is accessible through the biquadratic coupling shown in Eqn. \ref{eqn:Potential_energy}. The challenge now is to find which modes couple most strongly to an IR active phonon with a frequency that is accessible in a modern ultrafast optical experiment. 

\onecolumngrid

\twocolumngrid
We explore this by calculating nonequilibrium phonon dispersion curves as a function of all IR phonons polarized along the [100], [110], and [111] crystallographic directions. The nonequilibrium phonon dispersion curves for the highest frequency (16.7 THz) IR-active phonon displaced along the [111] direction, which we will use as a representative example, are shown in Fig. \ref{fig:KTaO3_phonon_dispersion}. As the amplitude of the IR phonon increases, we see a branch split at the R-point, with one mode softening and becoming unstable, that is, its frequency becomes imaginary (or its force constant becomes negative). Inspection of the displacement pattern and symmetry of this mode shows that it is an out-of-phase octahedral tilt (transforming as the irreducible representation R$_4^+$) about the [111] axis (Region II of Fig. \ref{fig:phase_diagram_expanded}). Octahedral tilts are common structural distortions in perovskites \cite{Goldschmidt1926,Lufaso2001} but are not present in KTaO$_3$ at any temperature. Here we show that a transient structural phase of KTaO$_3$ involving octahedral tilts can be induced by biquadratic coupling to an optically excited IR-active phonon. 

Since the mode that freezes in is at the Brillouin zone edge and has wave vector $\mathbf{q}_R = \frac{2\pi}{a}\left(\frac{1}{2},\frac{1}{2},\frac{1}{2}\right)$ (where $a$ is the cubic lattice parameter) the unit cell of the cubic phase is doubled to accommodate the octahedral tilt pattern. The transient unit cell lattice parameters ($\vec{a}'_i$) are given by $\vec{a}'_1 = \vec{a}_2 + \vec{a}_3$, $\vec{a}'_2 = \vec{a}_1 + \vec{a}_3$, and $\vec{a}'_3 = \vec{a}_1 + \vec{a}_2$, where the unprimed quantities denote the equilibrium lattice parameters and $\vec{a}_1 = \vec{a}_2 = \vec{a}_3$ since KTaO$_3$ is cubic at equilibrium. This shows that light-induced structural phases can change the translational symmetry of crystals on ultrafast timescales. In Sec. \ref{sec:polarization} we show that the changes to point group and translational symmetry are polarization dependent, allowing for unprecedented control of crystalline symmetry.

Using the DFT-derived parameters in Table \ref{tab:KTaO3_parameters} we can estimate the critical electric field (Eqn. \ref{eqn:crit_field}) needed to enter Region II of the phase diagram and induce a new structural phase in cubic KTaO$_3$. For a 500 fs Gaussian electric field pulse with 16.7 THz carrier frequency, we find a threshold field of 4.8 MV/cm. We note that this value ignores Fresnel reflection, higher-order anharmonicities, and damping.

\begin{center}
\begin{table}[h]
\caption{Characteristics of the high-frequency IR mode ($Q_{IR}$) and biquadratically coupled $R$-point modes ($Q_{\vec{q}}$) in the cubic phase of KTaO$_3$ from our DFT calculations needed to realize dynamical regimes in Region II and Region III. The IR mode is polarized along [111] in Region II and along [100] in Region III. In Region III, the excited IR mode and $Q_{\vec{q}}$ are very strongly coupled and $D_{IR,\vec{q}}$ varies from 10-20 eV/\AA$^4$, depending on the precise amplitudes of $Q_{IR}$ and $Q_{\vec{q}}$ used to calculate it. We used $D_{IR,\vec{q}}$ = 10 eV/\AA$^4$ for our dynamical simulations. The units of reduced mass are in atomic mass units, u.}

    \label{tab:KTaO3_parameters}
    \centering

{\setlength{\tabcolsep}{.45em}
    \begin{tabular}{l c c  r  r  r  }
    \hline\hline
    \multicolumn{3}{c}{Quantity} & \multicolumn{1}{c}{$Q_{IR}$} & \multicolumn{2}{c}{$Q_{\vec{q}}$} \\ \cline{5-6}
    & & &  & \multicolumn{1}{c}{Reg II} & \multicolumn{1}{c}{Reg III\rule{0pt}{9pt}}  \\ \hline 
        Frequency & $f$        &        [THz]       & 16.73 &  3.98 & 15.19  \rule{0pt}{10pt} \\ 
        Force constant & $K$        &  [eV/\AA$^2$] & 18.36 &  1.04  & 15.16 \rule{0pt}{10pt}\\ 
        Reduced mass & $M$           &       [u] & 16.03 & 16.00  & 16.06 \rule{0pt}{10pt}\\ 
    Mode-effective  \rule{0pt}{10pt}   & \multirow{2}{*}{$\tilde{Z}^*$} & \multirow{2}{*}{[e$^-$] } & \multirow{2}{*}{13.03} & \multicolumn{1}{c}{\multirow{2}{*}{-}} & \multicolumn{1}{c}{\multirow{2}{*}{-}} \\
     \multicolumn{1}{c}{charge} &                   &                   &                   &                   &                   \\
 Biquadratic  \rule{0pt}{10pt}& \multirow{2}{*}{$D_{IR,\vec{q}}$} & \multirow{2}{*}{ [eV/\AA$^4$]} & \multicolumn{1}{c}{\multirow{2}{*}{-}} & \multirow{2}{*}{-1.12} & \multirow{2}{*}{ 10.00} \\
 \multicolumn{1}{c}{coupling} &                   &                   &                   &                   &   \\   \hline\hline

    \end{tabular}}

\end{table}
\end{center}

We explore the dynamics for the excitation of the 16.7 THz IR active phonon polarized along the [111] direction as a function of peak electric field in the Region II panel inset of Fig. \ref{fig:phase_diagram_expanded}. The frequency ratio ($\sqrt{\delta}=0.24$) of the two phonons involved places us on the gray arrow originating on the vertical axis in the phase diagram and pointing into the gray region. A peak electric field of 3.7 MV/cm ($\epsilon = -0.033$) is not strong enough to overcome the intrinsic restoring force such that any transient motion imparted on $Q_{\vec{q}}$ is damped away, hence the amplitude of $Q_{\vec{q}}$ is zero. That is, the threshold from Region I to Region II is not crossed. By increasing the peak electric field to 5.3 MV/cm ($\epsilon=-0.066$), just beyond the critical field, a new energy minimum develops for $Q_{\vec{q}}$ at $Q^*_{\vec{q},+}$ (Eq. \ref{eqn:new_minima}). Because the critical condition has just been reached, the effective potential is shallow and the new resonant frequency $\omega^*$ (Eq. \ref{eqn:new_freq}) is small. Increasing the peak electric field further to 6.4 MV/cm ($\epsilon=-0.099$) further establishes the Region II behavior. The renormalized frequency $\omega^*$ has increased as expected, the location of the new energy minimum ($Q^*_{\vec{q},+}$) is at a larger amplitude, and the enhanced growth rate ($\mu_{II,+}$) towards the new phase is visually apparent. This is a consequence of the field-dependent growth rate in Eqn. \ref{eqn:region_II_growth}, giving access to the new structural phase $\approx 0.5$ ps earlier compared to the 5.3 MV/cm peak electric field. We expect the threshold values of $E_0 \tau$ (Eqn. \ref{eqn:crit_field}) to be accessible with current mid-IR laser sources at 16.7 THz \cite{Sell2008, Knorr2017}.

\subsubsection{Region III -- Parametric Oscillation}\label{sec:region_III}

In the previous section, driving an IR-active phonon condensed a single phonon, $Q_{\vec{q}}$, transiently altering the crystal symmetry. In complex crystals there are many modes of arbitrary wave vectors, all of which are accessible to the biquadratic coupling. As a result, we expect the dynamical response of the crystal to resonant IR-phonon drive to be rich with complex dynamical motion on many length scales involving many phonons. In this section we show that parametrically driven oscillatory motion of other $Q_{\vec{q}}$'s is also expected, adding a complexity of \emph{unexplored} time scales and microscopic motion to the dynamical response of the crystal.

We highlight the parametric oscillation of $Q_{\vec{q}}$ through the optical excitation of an IR-active phonon in the upper right of the phase diagram (positive $\epsilon$, blue region). Near $\epsilon=0$, this region grows from $\delta = 1$, the oscillatory component of the driver is twice the fundamental frequency of $Q_{\vec{q}}$ (cosine term in Eqn. \ref{eqn:hill_matrix}). This suggests that Region III describes parametric oscillation of $Q_{\vec{q}}$. We therefore expect rapidly growing oscillatory motion of $Q_{\vec{q}}$ in this region. Derivation of an approximate phase boundary, exponential growth rate, and the effect of intrinsic damping are included in Appendix \ref{app:Parametric}, following standard approaches to solutions in dynamical systems \cite{jose_1998}. 


\begin{center}
\begin{table}[h]
\caption{Characteristics of the IR mode ($Q_{IR}$) and biquadratically coupled $R$-point mode ($Q_{\vec{q}}$) in the cubic phase of SrTiO$_3$ from our DFT calculations needed to realize dynamical regimes in Region VI and Region V. The IR mode is polarized along [110] in Region VI and along [001] in Region V. The units of reduced mass are in atomic mass units, u. }

    \label{tab:SrTiO3_parameters}
    \centering
    
{\setlength{\tabcolsep}{.45em}
    \begin{tabular}{l c c  r  r  r  }
    \hline\hline
    \multicolumn{3}{c}{Quantity} & \multicolumn{1}{c}{$Q_{IR}$} & \multicolumn{2}{c}{$Q_{\vec{q}}$} \\ \cline{5-6}
    & & &  & \multicolumn{1}{c}{Reg IV} & \multicolumn{1}{c}{Reg V\rule{0pt}{9pt}}  \\ \hline 
        Frequency & $f$        &        [THz]       & 5.15 &  2.68$i$ & 2.68$i$  \rule{0pt}{10pt} \\ 
        Force constant & $K$        &  [eV/\AA$^2$] & 6.25 &  -0.47  & -0.47 \rule{0pt}{10pt}\\ 
        Reduced mass & $M$           &       [u] & 57.60 &  16.00  & 16.00 \rule{0pt}{10pt}\\ 
    Mode-effective  \rule{0pt}{10pt}   & \multirow{2}{*}{$\tilde{Z}^*$} & \multirow{2}{*}{[e$^-$] } & \multirow{2}{*}{11.90} & \multicolumn{1}{c}{\multirow{2}{*}{-}} & \multicolumn{1}{c}{\multirow{2}{*}{-}} \\
    \multicolumn{1}{c}{charge} &                   &                   &                   &                   &                   \\
 Biquadratic  \rule{0pt}{10pt}& \multirow{2}{*}{$D_{IR,\vec{q}}$} & \multirow{2}{*}{ [eV/\AA$^4$]} & \multicolumn{1}{c}{\multirow{2}{*}{-}} & \multirow{2}{*}{-0.29} & \multirow{2}{*}{ 0.85} \\
 \multicolumn{1}{c}{coupling} &                   &                   &                   &                   &   \\   \hline\hline

    \end{tabular}}
\end{table}
\end{center}




To give a quantitative example of the parametric oscillation process, we return to KTaO$_3$, and look for positive biquadratic coupling. We identify strong positive biquadratic coupling between the 16.7 THz IR phonon, now polarized along the [100] direction, and a phonon at the $M$-point of the Brillouin zone that transforms like the irreducible representation $M_3^-$ with frequency 15.2 THz (Fig. \ref{fig:phase_diagram_expanded}, Region III Structural Changes); the parameters we used for our dynamical simulations, calculated from first principles, are shown in Table \ref{tab:KTaO3_parameters}. The frequency ratio ($\sqrt{\delta}=0.91$) of the two phonons places us on the blue arrow originating from the vertical axis of the phase diagram, near the phase boundary. At this frequency ratio, parametric oscillation of the M$_3^-$ phonon is expected when $0.11<\epsilon<0.34$ (Eqn. \ref{eqn:parametric_boundary}) corresponding to peak electric fields between 2.3 MV/cm $<E_0<$ 4.0 MV/cm. This is consistent with the simulated dynamics in Fig. \ref{fig:phase_diagram_expanded}, Region III, which we will now discuss. 

For $ E_0 = 2.2 $ MV/cm ($\epsilon = 0.10 $) the amplitude of $Q_{\vec{q}}$ appears identical to the horizontal axis in the structural dynamics panel of Fig. \ref{fig:phase_diagram_expanded}, Region III. That is, the dynamics are those of Region I.

For $ E_0 = 3.3 $ MV/cm ($\epsilon = 0.23 $) we see large amplification of the $M_3^-$ phonon which grows over $\approx 1$ ps. The large amplification of $Q_{\vec{q}}$ is due to the oscillation of its effective potential through the $Q_{IR}$. This parametric oscillation mechanism is shown schematically in Fig. \ref{fig:phase_diagram_expanded}, Region III. The effective potential oscillates at a frequency of $2\omega_{IR}$ as a result of the IR phonon drive, stiffening the potential through the positive $D_{IR,\vec{q}}$ (negative $D_{IR,\vec{q}}$ would soften the potential). A half-cycle of $Q_{\vec{q}}$ motion is shown in four steps with $Q_{\vec{q}}$ starting at the apex of a cycle. During the first step (1), $Q_{\vec{q}}$ falls toward the static equilibrium point as the potential softens to its time-averaged value (Eqn. \ref{eqn:ZE_Force_constant}). In the second step (2), $Q_{\vec{q}}$'s kinetic energy allows it to move beyond the equilibrium point as the potential continues to soften to its undriven value. In the third step (3), $Q_{\vec{q}}$ begins losing kinetic energy while the potential stiffens back to the time-averaged value. In the final step (4), $Q_{\vec{q}}$ reaches the height of its half-cycle motion as the potential stiffens back to its apex. $Q_{IR}$'s effect on the potential increases the amplitude of $Q_{\vec{q}}$ with each half-cycle of motion, as expected in a parametric oscillation process. The growth in $Q_{\vec{q}}$ then decays back to zero amplitude due to the ``back action" on $Q_{IR}$. That is, $Q_{IR}$ is transiently driven with a finite amount of energy, as $Q_{\vec{q}}$ grows in amplitude it must gain energy from $Q_{IR}$, thereby decreasing the overall amplitude of the IR mode. Additionally, as the amplitude of $Q_{\vec{q}}$ grows, the frequency of $Q_{IR}$ will be modified via the biquadratic coupling, tuning the frequency ratio $\sqrt{\delta}$ away from Region III (see the $2 D_{IR,\vec{q}} Q_{\vec{q}}^2$ term in Eqn. \ref{eqn:EOM}). This suggests that other changes in the IR phonon frequency or amplitude, \textit{e.g.} through other anharmonic couplings, have a similar detrimental effect on the parametric oscillation process. 

For $ E_0 = 4.4 $ MV/cm ($\epsilon = 0.41 $) we are again in a trivial region of the phase diagram where excitation of $Q_{IR}$ has negligible effect on the amplitude of $Q_{\vec{q}}$.  This is because the effective frequency of $Q_{\vec{q}}$ (due to $Q_{IR}$) is driven out of sync with the oscillation in its potential energy landscape. This conversely explains why amplification was not seen for the $ E_0 = 2.2 $ MV/cm case, the effective frequency of $Q_{\vec{q}}$ was not driven high enough to sync up with $Q_{IR}$.

Since the parametric oscillation process is general, we expect \emph{many} modes to parametrically oscillate for a large enough drive, the effects of which we expect will be seen in structure factor analysis of diffuse scattering following IR excitation. That being said, parametric oscillation processes and the phase boundary between region I and region III are sensitive to damping (Appendix \ref{app:Parametric}). As a result, the experimental observation of parametric oscillation between the modes used in this illustrative example, even with the large coupling between them, may be hindered by damping.

We are unaware of any experimental work showing parametric oscillation through the anharmonic lattice potential in the nonlinear phononics literature. A recent work proposed parametric oscillation of an IR-active phonon through nonlinear (in $Q_{IR}$) contributions to the polarizability as an explanation for IR-resonant enhancement of the reflectivity in the reststrahlen band in SiC \cite{Cartella2018}; we have ignored the nonlinear polarizability contribution to the dynamics in this work, although this would certainly be an interesting path to pursue for future work.

What happens in crystals where a structural mode is already present in the equilibrium phase? We now consider the lower half of the phase diagram (imaginary $\sqrt{\delta} $), and look at Region IV to show that the conventional nonlinear phononics response may be activated by the condensation of a mode responsible for the structural phase transition.

\subsubsection{Region IV -- Conventional Nonlinear Phononics}\label{sec:region_IV}
We now briefly consider Region IV in the lower half of the phase diagram in Fig. \ref{fig:phase_diagram_expanded}. This region corresponds to the conventional nonlinear phononics effect, which has been the focus of many previous studies \cite{rini07,forst11,mankowsky14,forst15,khalsa18,Disa2021}. 

For this region, and in the lower part of the phase diagram generally, it is helpful to consider some (possibly virtual) high-symmetry reference phase of the material of interest. In this phase, $Q_{\vec{q}}$ has a negative force constant ($K_{\vec{q}} < 0$) and $\sqrt{\delta}$ is imaginary. The Floquet analysis predicts exponential growth of $Q_{\vec{q}}$ even in the absence of an excited IR phonon ($\epsilon = 0$) because $Q_{\vec{q}}$ is independently `unstable' and induces a non-transient structural phase transition. The symmetry of the reference phase is lowered by $Q_{\vec{q}}$ such that $Q_{\vec{q}}$ becomes fully symmetric (transforms like the identical representation) in the new structural phase. It is therefore Raman-active about its new minimum and described by a new mode $Q_R$ (that is $Q_{\vec{q}} \rightarrow Q_{\vec{q},\pm}^*\left(0\right) + Q_{R}$). The biquadratic coupling term between an IR-active mode and $Q_{\vec{q}}$ in the high-symmetry phase transforms into two lower-order terms describing a renormalization of the IR force constant of the form $K_{IR} \rightarrow K_{IR} + \left(2 D_{IR,\vec{q}} Q_{\vec{q},\pm}^* \left(0\right)^2 \right)$, and a new linear-quadratic coupling is created of the form $A Q_{IR}^2 Q_R $ (where $A =  2 D_{IR,\vec{q}} Q_{\vec{q},\pm}^*$(0)). 

When an IR phonon is excited in the low-symmetry phase it will exert a unidirectional force on other modes coupled to it through the $A Q_{IR}^2 Q_R $ term and the $Q_R$ modes will be unidirectionally displaced from their equilibrium amplitudes. The sign of the biquadratic coupling parameter $D_{IR,\vec{q}}$ dictates the direction in which $Q_R$ will be displaced: for $D_{IR,\vec{q}} < 0$ (Fig. \ref{fig:phase_diagram_expanded}, left side of Region IV), $Q_R$ is pushed away from the saddle point (the amplitude of $Q_R$ increases compared to its equilibrium value), whereas for $ D_{IR,\vec{q}} > 0$, $Q_{R}$ is pushed towards the saddle point (the amplitude of $Q_R$ \emph{decreases} relative to its equilibrium value). Since $Q_{R}$ is fully symmetric as mentioned above, unidirectional displacements in this region do not change the crystal point group or translational symmetry \cite{note4}. We will return to this point in Region V, where symmetry can be restored by driving $Q_{IR}$.

We focus our theoretical development of this region on SrTiO$_3$, which has been studied in several recent nonlinear phononics experiments \cite{Nova19,Li2019}. Our development here is intended to demonstrate conceptual features of the Floquet phase diagram, with a detailed description saved for a future publication.

To explore the dynamics and connect with the theoretical development of Sections \ref{sec:simple_model} and \ref{sec:Floquet}, we first calculate parameters for Eqn. \ref{eqn:Potential_energy} by considering an IR-active phonon at 5.15 THz in the cubic phase of SrTiO$_3$, which we take as our high-symmetry reference phase (see Table \ref{tab:SrTiO3_parameters}). We again consider coupling to a mode with $R_4^+$ symmetry, which corresponds to an out-of-phase tilt of the TiO$_6$ octahedra \cite{Fleury1968}. As mentioned above, this mode drives a structural phase transition in SrTiO$_3$ at about 110 K to a phase with $I4/mcm$ symmetry. Hence, in the cubic phase at 0 K this octahedral tilt mode has an imaginary frequency, 2.68$i$ THz. For this particular mode pairing, $\sqrt{\delta}=0.52i$, which places us in a region represented by the purple arrow originating from the vertical axis in Fig. \ref{fig:phase_diagram_expanded}, Region IV. 

We perform our simulations in a structural phase in which the octahedral tilt mode has frozen in to the cubic structure to produce the low-symmetry (low temperature) tetragonal $I4/mcm$ phase. In this phase, the octahedral tilt mode has an equilibrium amplitude of -72 pm from our DFT calculations and we denote it as $Q_R$ (the negative amplitude signifies the left minimum of the double-well potential in Fig. \ref{fig:phase_diagram_expanded} Region IV).
The triply degenerate IR phonon at 5.15 THz in the cubic phase is split in the tetragonal phase into a mode polarized along the axis about which the octahedra rotate (5.54 THz from our calculations), and a doubly degenerate mode polarized along the in-plane direction, perpendicular to the axis about which the octahedra rotate (5.04 THz). In this section we focus on this latter set of modes.  

Excitation of an IR phonon polarized along the in-plane direction quasi-statically displaces $Q_R$ away from the saddle point shown in the Region IV inset in Figure \ref{fig:phase_diagram_expanded}; that is, the amplitude of $Q_R$ transiently increases. As the peak electric field increases from 5 MV/cm to 7 MV/cm and 9 MV/cm, the amplitude increases to -85 pm, -97 pm and -112 pm (corresponding to an $\epsilon$ of -0.11, -0.22, -0.37).

In this section, we have shown that the conventional nonlinear phononics effect can be incorporated into a more general parametric amplification framework that is enabled by biquadratic coupling between modes in a proximal higher symmetry parent phase that is either real or virtual. In the next section, we go one step further and show that higher symmetry phases may be stabilized through IR-active phonon drive.

\subsubsection{Region V -- (Re)Introducing Symmetry}\label{sec:region_V}

In this final section, we show that crystal phases of proximal high-symmetry parent phases can be stabilized in the transient response to IR-phonon drive, (re)introducing symmetry elements into the transient crystal structure. 

For this region it is again helpful to consider some high-symmetry reference phase of the material of interest, as was done in the discussion of Region IV. In this reference phase, $Q_{\vec{q}}$ has a negative force constant ($K_{\vec{q}} < 0$) and $\sqrt{\delta}$ is imaginary -- the high-symmetry reference structure is a saddle-point of the energy. With $D_{IR,\vec{q}} > 0$ and a sufficient drive $\epsilon$, Floquet theory predicts exponential decay of $Q_{\vec{q}}$. That is, Floquet theory predicts the high-symmetry reference phase is stable for large enough IR-active phonon drive, essentially stabilizing a saddle-point of the energy landscape. The negative $\tilde{K}_{\vec{q}}$ is overcome and made positive by the driven oscillating potential through the large amplitude IR-active phonon motion and biquadratic coupling. This is the phonon counterpart to the classic rigid pendulum problem where driving its pivot point periodically can stabilize the inverted solution \cite{Stephenson1908,Kapitsa_1951}. 

 

Mathematically, the results are identical to those found in Sec. \ref{sec:region_II}, but both $\delta$ and $\epsilon$ have changed sign. Rather than making $\tilde{K}_{\vec{q}}$ negative by driving the IR mode, $\tilde{K}_{\vec{q}}$ starts negative and is driven to a positive value, collapsing the time-averaged double-well back to a stable single-well. The phase boundary is therefore defined by, in analogy to the discussion below Eqn. \ref{eqn:region_II_approx_growth},  $\epsilon=|\delta|$. In this way the minima associated with $Q_{\vec{q},\pm}^*\left(0\right)$ have moved to zero, restoring a higher-symmetry crystal configuration by reintroducing the symmetry elements of $Q_{\vec{q}}$.

To illustrate this point, we focus on the out-of-plane polarized IR mode in SrTiO$_3$ at 5.54 THz mentioned in Sec. \ref{sec:region_IV}. In contrast to the in-plane excitation,  the sign of the biquadratic coupling is positive and therefore $\epsilon>0$ (Table \ref{tab:SrTiO3_parameters}). Exciting this phonon pushes the octahedral tilts ($Q_R$) from their starting amplitude of -72 pm towards zero amplitude. For $\epsilon=0.21$ ($E_0=4.0$ MV/cm), the response is still of the conventional nonlinear phononics type. That is, the amplitude of $Q_R$ changes but the point group and translation symmetry of the crystal is preserved. For $\epsilon = 0.65 $ ($E_0=7.0$ MV/cm), we crossover the phase boundary from Region IV to Region V, and the point group and translational symmetry elements are reintroduced so that the average transient structure appears cubic ($Pm\bar{3}m$, space group \#221). 

In the transient response, as $Q_{IR}$ dissipates energy it eventually becomes unable to sustain $Q_R$ about the cubic structure. When this happens $Q_R$ will fall back into either of the double-well minima (Fig. \ref{fig:phase_diagram_expanded}, Region V). We expect that in experiments this process will depend sensitively on details of the pulse characteristics, damping, initial conditions, and the boundary of the illuminated region of the crystal \cite{mankowsky17,Mertelj2019,Abalmasov2020}. This suggests that with these approximations \emph{deterministic} switching from one double-well minimum to another is not possible by this pathway. This was pointed out in a recent study where a multi-pulse sequence was needed to switch the polarization of KNbO$_3$ \cite{Chen2022}, and may partly explain the lack of switching and transient recovery of polarization seen in an IR phonon pumped experiment in LiNbO$_3$ \cite{mankowsky17} (as pointed out in \cite{Mertelj2019, Abalmasov2020}). 

For large enough drive, another phase boundary is crossed to the parametric oscillation regime (Region III). In the transient response, in order to cross to the parametric oscillation regime, the trajectory must pass through Region V. That is, the transient response will first be stabilized in the high-symmetry structure before the mode parametrically oscillates. This is shown in the simulated dynamics of Fig. \ref{fig:phase_diagram_expanded}, Region V for $\epsilon = 0.93 $ and $E_0=8.4$ MV/cm. Increasing $\epsilon$ further will traverse an alternating series of phase boundaries of reintroducing symmetry and parametric oscillation regimes, though we expect this to be largely inaccessible in experiment due to the presence of other anharmonicity or destruction/melting of the crystal.


\subsection{Mode- and polarization-selective response}\label{sec:polarization}

In the development of the work presented so far, it is implicit that the response is frequency dependent. That is, the choice of IR phonon to excite, along with its polarization, may strongly affect the resulting transient structural phase transition. In this section, we illustrate the mode- and polarization-selective features of the lattice response by focusing on Region II of the phase diagram for KTaO$_3$ and SrTiO$_3$. For KTaO$_3$ we can estimate the critical fields needed to induce new structural phases since the cubic phase is stable in DFT. For SrTiO$_3$, since the cubic phase is unstable in DFT (there are phonon modes with imaginary frequencies), we report only the new structural phases induced, which we expect to be relevant to experiments just above the structural phase transition temperature of 110 K. We note that a general exploration of the polarization dependence for all regions is possible with \emph{ab initio} techniques, but quite computationally expensive due to the enormous number of anharmonic pathways allowed. This is particularly true for Region III, where phonons at all frequencies and all wave vectors may be involved. 

In both KTaO$_3$ and SrTiO$_3$, and in perovskite materials in general, the most common structural instabilities are associated with the R$_4^+$ and M$_3^+$ zone-edge phonons, that is, these modes often appear with imaginary frequencies in the cubic phase (readily calculated using DFT) and often drive structural phase transitions. The displacement patterns for these phonons represent out-of-phase (R$_4^+$), and in-phase (M$_3^+$) octahedral tilts of the TaO$_6$ or TiO$_6$ octahedra about the cubic crystallographic axes.

It is convenient to introduce a notation favored in the complex oxide community, after Glazer \cite{Glazer1972Notation1,Glazer1975Notation2}, which describes the in-phase (+) and out-of-phase (-) octahedral tilts as a list about the cubic-crystallographic axes. In the example described in Fig. \ref{fig:phase_diagram_expanded} for Region II, the equilibrium state of KTaO$_3$ is described by the label $a^0a^0a^0$ indicating that the $a$, $b$, and $c$ lattice constants are all equal and that there are no octahedral tilts about any crystallographic axis. The octahedral tilt pattern induced in Region II (associated with the $R_4^+[111]$ phonon of the cubic phase) following excitation of an IR phonon polarized along the [111]-direction is labeled $a^-a^-a^-$, signifying that the octahedral tilts about each crystallographic axes are all out-of-phase with respect to each other but are of the same amplitude. The octahedral tilt patterns in KTaO$_3$ associated with transient structural phase transitions following excitation of various IR-active phonons polarized along the [100], [110], and [111] directions are shown in Table \ref{tab:KTaO3_instabilities} (Table \ref{tab:SrTiO3_instabilities} for SrTiO$_3$). The entries shown represent the first modes that develop negative force constants (imaginary frequencies) with respect to an increase in amplitude of a given IR mode (see Fig. \ref{fig:KTaO3_phonon_dispersion}). We expect that polarization directions between the principal crystallographic axes will give octahedral tilt patterns between those listed in Table \ref{tab:KTaO3_instabilities}, \textit{e.g.} for a 16.7 THz pulse polarized between [111] and [110] we expect an octahedral tilt pattern of $a^-a^-b^-$, which corresponds to the $C2/c$ ($C_{2h}$) space group. Note that for the [100] direction, the two directions of $Q_{\vec{q}}$ shown in Tables \ref{tab:KTaO3_instabilities} \& \ref{tab:SrTiO3_instabilities} have the same energy decrease in DFT, so either direction, or both directions may be seen in experiment.

Depending on the polarization direction, although the octahedral tilts dominate the induced $Q_{\vec{q}}$ structural change, other structural distortions may be present, including strain (Supplementary Sec. \ref{sec:Strain}). We find that other structural distortions associated with A-site and B-site motion and deformation of the oxygen octahedra tend to be small compared to the octahedral tilt components of the motion (Supplementary Sec. \ref{sec:eigenmodes}). 

We speculate that the strong coupling between IR phonons and octahedral tilt modes is a consequence of the geometric network of bonds in both KTaO$_3$ and SrTiO$_3$, and is likely general to perovskites. To illustrate this, consider the 16.7 THz IR phonon polarized along [100] direction in KTaO$_3$, where at the critical field, the shortest time-averaged Ta-O bond is along the polarization direction and its length has decreased by $\approx \pm$10\% ($\approx \pm20$ pm). This shortest Ta-O bond is energetically unfavorable. Rotating the octahedra by displacing the $A_5$ symmetry $Q_{\vec{q}}$ mode (Table \ref{tab:KTaO3_instabilities}) accommodates this unfavorable condition by increasing the length of this Ta-O bond towards its equilibrium value (Supplementary Fig. \ref{fig:Bondlength})

\begin{widetext}

\begin{table}[ht!]
\caption{
Induced instabilities (Region II) in KTaO$_3$ for resonant excitation of  IR-active phonons polarized along different directions from our DFT calculations. Within a half cycle of the IR-active phonon, the space group symmetry is lowered (third column). As the IR-active phonon rings,  $Q_{\vec{q}}$ condenses into the crystal, further lowering the space group symmetry (fifth column), altering the point group and translational symmetry of the crystal.  $Q_{\vec{q}}$, labeled by the irreducible representation (Irrep) in the space group induced by the IR-active phonon excitation (fourth column), is primarily associated with TaO$_6$ octahedral tilt patterns (sixth column), but may also include other subtle distortions (Supplementary Sec. \ref{sec:eigenmodes}). The space group induced by the octahedral tilts alone is given in the seventh column. The critical electric field needed to condense $Q_{\vec{q}}$ for a 500 fs duration Gaussian electric field pulse is given in the last column. For the 5.5 THz IR phonon polarized along the [110] direction, our calculations suggest that higher-order lattice anharmonicity may be needed to describe the condensation of $Q_{\vec{q}}$.
}
\label{tab:KTaO3_instabilities}

{\setlength{\tabcolsep}{.63em}
\begin{tabular}{r c c c c c c r}
\hline \hline
\multicolumn{1}{c}{\begin{tabular}[c]{@{}c@{}}Frequency\\ {[}THz{]}\end{tabular}} & \begin{tabular}[c]{@{}c@{}}Polarization\\ Direction\end{tabular} & \begin{tabular}[c]{@{}c@{}}Space group\\ induced by \\ $Q_{IR}$\end{tabular} & \begin{tabular}[c]{@{}c@{}}Irrep of $Q_{\vec{q}}$ in\\ space group\\ induced by $Q_{IR}$\end{tabular} & \begin{tabular}[c]{@{}c@{}}Space group\\ induced by\\ $Q_{IR} + Q_{\vec{q}}$\end{tabular} & \begin{tabular}[c]{@{}c@{}}Octahedral\\ tilt pattern\end{tabular} & \begin{tabular}[c]{@{}c@{}}Space group\\ induced by\\ octahedral tilt\end{tabular} & \multicolumn{1}{c}{\begin{tabular}[c]{@{}c@{}}Critical field\\ {[}MV/cm{]}\end{tabular}} \\ \hline
\rule{0pt}{12pt} 
16.7 & \multirow{3}{*}{{[}100{]}} & \multirow{3}{*}{P4mm (C$_{4v}$)} & \multirow{3}{*}{\begin{tabular}[c]{@{}c@{}}A$_5$(a,0)\\ or\\ A$_5$(a,a)\end{tabular}} & \multirow{3}{*}{\begin{tabular}[c]{@{}c@{}}Ima2 (C$_{2v}$)\\ or\\ Fmm2 (C$_{2v}$)\end{tabular}} & \multirow{3}{*}{\begin{tabular}[c]{@{}c@{}}$a^0b^-b^-$\\ or\\ $a^0a^0b^-$\end{tabular}} & \multirow{3}{*}{\begin{tabular}[c]{@{}c@{}}Imma (D$_{2h}$)\\ or\\ I4/mcm (D$_{4h}$)\end{tabular}} & 4.5
 \\
5.5 &  &  &  &  &  &  & \multicolumn{1}{r}{18.0} \\
3.3 &  &  &  &  &  &  & \multicolumn{1}{r}{1.8} \\ 
\rule{0pt}{25pt} 
16.7 & \multirow{3}{*}{{[}110{]}} & \multirow{3}{*}{Amm2 (C$_{2v}$)} & T$_4$ & Ima2 (C$_{2v}$) & $a^0a^0b^-$ & I4/mcm (D$_{4h}$) & 4.6
 \\
5.5 &  &  & Y$_4$ & Pmc2$_1$ (C$_{2v}$) & $a^0a^0b^+$ & P4/mbm (D$_{4h}$) & 11.6 \\
3.3 &  &  & T$_4$ & Ima2 (C$_{2v}$) & $a^0a^0b^-$ & I4/mcm (D$_{4h}$) & 2.3
 \\ 
\rule{0pt}{25pt} 
16.7 & \multirow{3}{*}{{[}111{]}} & \multirow{3}{*}{R3m (C$_{3v}$)} & \multirow{3}{*}{T$_2$} & \multirow{3}{*}{R3m (C$_{3v}$)} & \multirow{3}{*}{$a^-a^-a^-$} & \multirow{3}{*}{R$\bar{3}$c (D$_{3d}$)} & 4.8
 \\
5.5 &  &  &  &  &  &  &  15.1 \\
3.3 &  &  &  &  &  &  &   2.8
\\ \hline \hline
\end{tabular}}
\end{table}

\begin{table}[ht!]
\caption{
Transiently induced structural phases (Region II) in SrTiO$_3$ for resonant excitation of  IR-active phonons polarized along different directions. Since the cubic phase of SrTiO$_3$ is dynamically unstable at 0 K in DFT, we only report the predicted structural phases, assuming that the critical field can be controlled by proximity to the 110 K phase transition temperature. Within a half cycle of the IR-active phonon, the space group symmetry is lowered (third column). As the excited IR-active phonon rings,  $Q_{\vec{q}}$ freezes into the crystal, further lowering the space group symmetry (fifth column), altering the point group and translational symmetry of the crystal.  $Q_{\vec{q}}$, labeled by the irreducible representation (Irrep) in the space group induced by the IR-active phonon excitation (fourth column), is primarily associated with TiO$_6$ octahedral tilt patterns (sixth column), but may also include other subtle distortions (Supplement Sec. \ref{sec:eigenmodes}). The space group induced by the octahedral tilts \emph{alone} is given in the seventh column.}

\label{tab:SrTiO3_instabilities}
{\setlength{\tabcolsep}{1.1em}
\begin{tabular}{rcccccc}
\hline \hline
\multicolumn{1}{c}{\begin{tabular}[c]{@{}c@{}}Frequency\\ {[}THz{]}\end{tabular}} & \begin{tabular}[c]{@{}c@{}}Polarization\\ Direction\end{tabular} & \begin{tabular}[c]{@{}c@{}}Space group\\ induced by \\ $Q_{IR}$\end{tabular} & \begin{tabular}[c]{@{}c@{}}Irrep of $Q_{\vec{q}}$ in\\ space group\\ induced by $Q_{IR}$\end{tabular} & \begin{tabular}[c]{@{}c@{}}Space group\\ induced by\\ $Q_{IR} + Q_{\vec{q}}$\end{tabular} & \begin{tabular}[c]{@{}c@{}}Octahedral\\ tilt pattern\end{tabular} & \begin{tabular}[c]{@{}c@{}}Space group\\ induced by\\ octahedral tilt\end{tabular} \\ \hline
\rule{0pt}{9pt} 
16.8 & \multirow{3}{*}{{[}100{]}} & \multirow{3}{*}{P4mm (C$_{4v}$)} & \multirow{3}{*}{\begin{tabular}[c]{@{}c@{}}A$_5$(a,0)\\ or\\ A$_5$(a,a)\end{tabular}} & \multirow{3}{*}{\begin{tabular}[c]{@{}c@{}}Ima2 (C$_{2v}$)\\ or\\ Fmm2 (C$_{2v}$)\end{tabular}} & \multirow{3}{*}{\begin{tabular}[c]{@{}c@{}}$a^0b^-b^-$\\ or\\ $a^0a^0b^-$\end{tabular}} & \multirow{3}{*}{\begin{tabular}[c]{@{}c@{}}Imma (D$_{2h}$)\\ or\\ I4/mcm (D$_{4h}$)\end{tabular}} \\
5.2 &  &  &  &  &  &  \\
1.9 &  &  &  &  &  &  \\ 
\rule{0pt}{25pt}
16.8 & \multirow{3}{*}{{[}110{]}} & \multirow{3}{*}{Amm2 (C$_{2v}$)} & \multirow{3}{*}{T$_4$} & \multirow{3}{*}{Ima2 (C$_{2v}$)} & \multirow{3}{*}{$a^0a^0b^-$} & \multirow{3}{*}{I4/mcm (D$_{4h}$)} \\
5.2 &  &  &  &  &  &  \\
1.9 &  &  &  &  &  &  \\ 
\rule{0pt}{25pt}
16.8 & \multirow{3}{*}{{[}111{]}} & \multirow{3}{*}{R3m (C$_{3v}$)} & \multirow{3}{*}{T$_2$} & \multirow{3}{*}{R3c (C$_{3v}$)} & \multirow{3}{*}{$a^-a^-a^-$} & \multirow{3}{*}{R$\bar{3}$c (D$_{3d}$)} \\
5.2 &  &  &  &  &  &  \\
1.9 &  &  &  &  &  &  \\ \hline \hline
\end{tabular}}
\end{table}
\end{widetext}

\section{Summary and Conclusions}
We have used a combination of first-principles DFT calculations and Floquet theory to develop a phase diagram depicting the various dynamical regimes accessible to materials given ultrafast optical excitation of an IR-active phonon biquadratically coupled to another mode at arbitrary wave vector. We have shown that crystal point group and translational symmetries may be introduced or removed via various mechanisms, depending on which dynamical regime is accessed in a given experiment. Our phase diagram is intended to frame theoretical and experimental work in the nonlinear phononics field where the transient response of the crystal may approach the Floquet regime, as justified by our dynamical simulations with parameters derived from first-principles calculations. Although we have ignored damping of the IR phonon in this work, we note that the inclusion of damping will quantitatively alter some of our results. That is, for a given IR-active phonon, larger peak electric fields and/or pulse durations may be needed to observe the desired effect.

As an example of point group and translational symmetry control, we have shown that in the perovskite KTaO$_3$, which is cubic at all temperatures at equilibrium, zone-edge octahedral tilts can be induced by coupling to an optically excited IR phonon, revealing a hidden structural phase. The explored polarization dependence of this phenomenon suggests that octahedral tilt modes (and potentially other kinds of structural distortion patterns) can be precisely controlled with light. We expect that the susceptibility of a given material to this kind of control is tied to the frequency of $Q_{\vec{q}}$, with lower frequency $\omega_{\vec{q}}$ modes more likely to induce transient structural phase transitions. Accompanying this lowering in symmetry, we expect other phonons to be amplified to large oscillatory motion through parametric oscillation. Resolving this motion experimentally will require measuring the structural response of the crystal on multiple timescales using, for example, ultrafast IR pump/diffuse X-ray scattering probes. Furthermore, our results suggest that continued development of tunable lasers spanning the 1 THz -- 20 THz spectral range will enable further exploration of the structural and functional response of crystals to light.

Finally, we note that in the development of the Floquet phase diagram in Fig. \ref{fig:phase_diagram_expanded} we have focused on coupling between the driven IR-active phonon and other phonons, however the form of Eqn. \ref{eqn:Potential_energy} is general. That is, replacing $Q_{\vec{q}}$ with an arbitrary order parameter describing, for example, magnetism or orbital order, preserves the symmetries of the model. The approach here is therefore general and can be applied to understand \emph{any} IR-phonon-driven phase change.

\acknowledgements
JZK and NAB were supported by the Department of Energy -- Office of Basic Energy Sciences under award DE-SC0019414. GK was supported by the Cornell Center for Materials Research with funding from the NSF MRSEC program (Grant No. DMR-1719875). Computational resources were provided by the Cornell Center for Advanced Computing.



\begin{appendix}
\section{$\epsilon$ Unit Conversion} \label{app:units}
The proper unit conversions are required in order for $\epsilon$ to be a dimensionless quantity. $\epsilon$, when simplified and written in terms of pulse characteristics and microscopic parameters

\begin{equation} \label{eqn:epsilon}
    \epsilon = \left(\frac{\eta}{\pi} \right) ^2   \frac{D_{IR,\vec{q}}   \tilde{Z}^{*2} \tau^2  E_{0}^2 }{ M_{\vec{q}} K_{IR}^2   } ,
\end{equation}

\noindent
where $\left(\frac{\eta}{\pi} \right) ^2 $ is a dimensionless coefficient and $\eta$ is defined by the pulse shape (see below Eqn. \ref{eqn:IR_approx}). Given the following choice of units for the other terms:  $D_{IR,\vec{q}}$ [eV/\AA$^4$], $\tilde{Z}^*$ [e$^-$], $\tau$ [ps], $E_0$ [MV/cm], $ M_{\vec{q}}$ [u (atomic mass units)], and $K_{IR}$ [eV/\AA$^2$]; a scale factor of 0.9648533 is needed.

\section{Constructing the Floquet Phase Diagram }\label{app:Floquet_machinery}
In order to construct the phase diagram for the dynamical response described by Eqn. \ref{eqn:hill} we recall the main results of Floquet theory \cite{magnus2004hill,Rand2018}. 

The fundamental solutions of Eqn. \ref{eqn:hill} are of the form, $x_i(\theta) = \lambda_i^{\theta / \pi} p_i(\theta)$ where ${p_i(\theta) = p_i(\theta+\pi) }$ shares the periodicity of $A(\theta)$ and $\lambda_i$ are the so-called \emph{characteristic multipliers}. This shows that solutions are generally periodic in time and may grow or decay exponentially. The $\lambda_i$ are found numerically by integrating Eqn. \ref{eqn:hill} over one period given a set of linearly independent initial conditions ${X(\theta_0) = \left( x_1(\theta_0),x_2(\theta_0),...,x_N(\theta_0) \right)}$. The transformation matrix $B = X(\theta_0)^{-1} X(\theta_0 + \pi)$ is diagonalized to find its eigenvalues -- the characteristic multipliers $\lambda_i$. Three scenarios are possible: $|\lambda_i| > 1$ corresponding to exponential growth, $|\lambda_i| < 1$ corresponding to exponential decay, and $|\lambda_i| = 1$ corresponding to stability of the $i^{th}$ solution. Therefore, $|\lambda_i| = 1$ define phase boundaries between regions of exponential growth and decay.

We numerically integrate Eqn. \ref{eqn:hill_matrix} from $\theta \in [0,\pi]$ for a mesh of $\delta$, $\epsilon$, and $\nu$ with initial conditions $x_1(0) = (1,0)$ and $x_2(0) = (0,1)$. $x_1(0)$ corresponds to a physical scenario where at $\theta=0$, $Q_{\vec{q}}$ is displaced but its velocity $\dot{Q}_{\vec{q}}$ is zero. Conversely, $x_2(0)$ corresponds to a scenario where at $\theta=0$, $Q_{\vec{q}}=0$ and the velocity is nonzero. The eigenvalues of the transformation matrix $B$ define the characteristic multipliers, which are analyzed to construct the phase boundaries ($\left|\lambda\right|=1$) and regions of exponential growth ($\left|\lambda\right|>1$) and decay ($\left|\lambda\right|<1$) in $Q_{\vec{q}}$.

\section{Parametric Oscillation Derivation}\label{app:Parametric}

In this Appendix, we derive expressions for the approximate phase boundary, an exponential growth rate, and peak growth rate including the effects of damping for parametric oscillation in Region III. We anticipate that these expressions will be useful in future experimental work exploring this phenomenon.

To accommodate the exponential growth predicted by the Floquet analysis and the expected periodic motion, we assume a solution of the form $Q_{\vec{q}} = A(\tau) cos(\tau) + B(\tau) sin(\tau)$. Inserting this \emph{ansatz} into Eqn. \ref{eqn:hill}, we find for the exponential growth parameters $A$ and $B$,

\begin{widetext}

\begin{equation} \label{eqn:parametric_matrix}
\begin{split}
    \frac{d}{d\tau }  \left( \begin{array}{cc}
    A \\
    B\\
    \end{array} \right) =
    \frac{1}{2}   \begin{bmatrix}
 0 & -1 + (\delta + \epsilon) - \epsilon/2 \\
1 - (\delta + \epsilon) - \epsilon/2 & 0 
\end{bmatrix}
    \left( \begin{array}{cc}
    A \\
    B\\
    \end{array} \right)
\end{split}
\end{equation}

\end{widetext}

\noindent
where $\ddot{A}$ and $\ddot{B}$ have been neglected in the spirit of the slowly varying envelope approximation \cite{Slowly_Varying_Envelop_Approx_1965}. We have also neglected high-harmonic terms and damping to find this form. Eqn. \ref{eqn:parametric_matrix} has exponential solutions with their growth rate found by solving for the eigenvalues of the matrix on the right-hand side. We find ${\mu_{V,\pm} = {\pm \frac{1}{2} \sqrt{ \left(\epsilon/2 \right)^2  - \left( 1 - \left( \delta + \epsilon \right) \right)^2 }}}$, where $\mu_{V,+}$ gives rise to exponential growth. The phase boundary is again identified by setting $\mu=0$. We find the following conditions for the phase boundary:

\begin{equation}\label{eqn:parametric_boundary}
\begin{split}
        \left( \frac{\epsilon}{2} \right)^2 \geq \left( 1 - \left( \delta + \epsilon \right) \right)^2\\
        \frac{2}{3}  \left( 1 - \delta \right) \leq  \epsilon  \leq 2 \left( 1 - \delta \right)     
\end{split}
\end{equation}

\noindent
Maximum exponential growth is found at $\epsilon^* = \frac{4}{3} \left( 1- \delta \right)$ with a growth rate of $\mu(\epsilon^*) = \frac{|1-\delta|}{2\sqrt{3}}$.

To account for the effect of damping we assume the standard result from damped oscillators $Q_{\vec{q}} \propto e^{\frac{\nu}{2} \tau}$ so that exponential growth is only expected when $\mu_{V,+}$ is greater than $\frac{\nu}{2}$. This alters Eqn. \ref{eqn:parametric_boundary} so that the phase boundary is defined by

\begin{equation}\label{eqn:parametric_boundary_damping}
\begin{split}
        \epsilon \geq \frac{4}{3}  \left( \left( 1 - \delta_{\nu} \right) -\frac{1}{2}\sqrt{\left(1 - \delta_{\nu}\right)^2 - 3\nu^2} \right) \\
        \epsilon \leq \frac{4}{3}  \left( \left( 1 - \delta_{\nu} \right) +\frac{1}{2}\sqrt{\left(1 - \delta_{\nu}\right)^2 - 3\nu^2} \right) 
\end{split}
\end{equation}

\noindent
Here $\delta_{\nu} = \delta - \nu^2/4$ accounts for the frequency shift imparted by the damping. This relation requires $|1-\delta_{\nu}| \ge \sqrt{3}\nu$ and $|\epsilon| \ge 2\nu$ for exponential growth. That is, there is a range of $\omega_{\vec{q}}$ near $\omega_{IR}$ that will not exhibit parametric oscillation and a larger drive $\epsilon$ is needed to overcome the damping for $\omega_{\vec{q}}$ outside this range. The decrease in size of Region III is taken up by Region I and Region V (Fig. \ref{fig:Damping}). The phase boundaries between Region V and Regions I \& IV represents the only phase boundaries in Fig. \ref{fig:phase_diagram_expanded} sensitive to damping of $Q_{\vec{q}}$.

\end{appendix}


\bibliography{citations.bib}

\begin{thebibliography}{10}

\bibitem{Goldschmidt1926}
V.~M. Goldschmidt, ``{Die Gesetze der Krystallochemie},'' {\em
  Naturwissenschaften}, vol.~14, pp.~477--485, May 1926.

\bibitem{Landau1965}
``29 - on the theory of phase transitions,'' in {\em Collected Papers of L.D.
  Landau} (D.~{TER HAAR}, ed.), pp.~193--216, Pergamon, 1965.

\bibitem{Shirane1974}
G.~Shirane, ``Neutron scattering studies of structural phase transitions at
  {Brookhaven},'' {\em Rev. Mod. Phys.}, vol.~46, pp.~437--449, July 1974.

\bibitem{Li2021}
W.~Li, X.~Qian, and J.~Li, ``Phase transitions in {2D} materials,'' {\em Nature
  Reviews Materials}, vol.~6, pp.~829--846, Sept. 2021.

\bibitem{toledano1987}
J.~Tolédano and P.~Tolédano, {\em The Landau Theory of Phase Transitions}.
\newblock World Scientific Publishing Company, 1987.

\bibitem{Borzi2007}
R.~A. Borzi, S.~A. Grigera, J.~Farrell, R.~S. Perry, S.~J.~S. Lister, S.~L.
  Lee, D.~A. Tennant, Y.~Maeno, and A.~P. Mackenzie, ``Formation of a {Nematic
  Fluid at High Fields} in {Sr$_3$Ru$_2$O$_7$},'' {\em Science}, vol.~315,
  pp.~214--217, Jan. 2007.

\bibitem{Gupta2021}
N.~K. Gupta, C.~McMahon, R.~Sutarto, T.~Shi, R.~Gong, H.~I. Wei, K.~M. Shen,
  F.~He, Q.~Ma, M.~Dragomir, B.~D. Gaulin, and D.~G. Hawthorn, ``Vanishing
  nematic order beyond the pseudogap phase in overdoped cuprate
  superconductors,'' {\em Proceedings of the National Academy of Sciences},
  vol.~118, p.~e2106881118, Aug. 2021.

\bibitem{Cao2018}
Y.~Cao, V.~Fatemi, S.~Fang, K.~Watanabe, T.~Taniguchi, E.~Kaxiras, and
  P.~Jarillo-Herrero, ``Unconventional superconductivity in magic-angle
  graphene superlattices,'' {\em Nature}, vol.~556, pp.~43--50, Mar. 2018.

\bibitem{Zhang_2012}
X.~Zhang, M.~Takahashi, K.~Takeuchi, and S.~Sakai, ``64 kbit
  {Ferroelectric-Gate-Transistor-Integrated NAND Flash Memory with 7.5 V
  Program and Long Data Retention},'' {\em Japanese Journal of Applied
  Physics}, vol.~51, p.~04DD01, Apr. 2012.

\bibitem{Jiao2020}
P.~Jiao, K.-J.~I. Egbe, Y.~Xie, A.~M.~Nazar, and A.~H. Alavi, ``{Piezoelectric
  Sensing Techniques in Structural Health Monitoring: A State-of-the-Art
  Review},'' {\em Sensors}, vol.~20, July 2020.

\bibitem{piezo1880}
P.~Curie and J.~Curie, ``Développement par compression de l'électricité
  polaire dans les cristaux hémièdres à faces inclinées,'' {\em Bulletin de
  Minéralogie}, vol.~3, no.~4, pp.~90--93, 1880.

\bibitem{Lippmann_1881}
G.~Lippmann, ``Principe de la conservation de
  l{\textquotesingle}{\'{e}}lectricit{\'{e}}, ou second principe de la
  th{\'{e}}orie des ph{\'{e}}nom{\`{e}}nes {\'{e}}lectriques,'' {\em Journal de
  Physique Th{\'{e}}orique et Appliqu{\'{e}}e}, vol.~10, no.~1, pp.~381--394,
  1881.

\bibitem{bousquet08}
E.~Bousquet, M.~Dawber, N.~Stucki, C.~Lichtensteiger, P.~Hermet, S.~Gariglio,
  J.-M. Triscone, and P.~Ghosez, ``Improper ferroelectricity in perovskite
  oxide artificial superlattices,'' {\em Nature}, vol.~452, pp.~732--736, Apr.
  2008.

\bibitem{MAH8}
T.~Fukushima, A.~Stroppa, S.~Picozzi, and J.~M. Perez-Mato, ``Large
  ferroelectric polarization in the new double perovskite {NaLaMnWO$_6$}
  induced by non-polar instabilities,'' {\em Phys. Chem. Chem. Phys.}, vol.~13,
  pp.~12186--12190, June 2011.

\bibitem{benedek11}
N.~A. Benedek and C.~J. Fennie, ``Hybrid improper ferroelectricity: A mechanism
  for controllable polarization-magnetization coupling,'' {\em Phys. Rev.
  Lett.}, vol.~106, p.~107204, Mar. 2011.

\bibitem{benedek12}
N.~A. Benedek, A.~T. Mulder, and C.~J. Fennie, ``Polar octahedral rotations: A
  path to new multifunctional materials,'' {\em Journal of Solid State
  Chemistry}, vol.~195, pp.~11--20, Nov. 2012.

\bibitem{benedek22}
N.~A. Benedek and M.~A. Hayward, ``{Hybrid Improper Ferroelectricity: A
  Theoretical, Computational, and Synthetic Perspective},'' {\em Annual Review
  of Materials Research}, vol.~52, pp.~331--355, July 2022.

\bibitem{rini07}
M.~Rini, R.~Tobey, N.~Dean, J.~Itatani, Y.~Tomioka, Y.~Tokura, R.~W.
  Schoenlein, and A.~Cavalleri, ``{Control of the electronic phase of a
  manganite by mode-selective vibrational excitation},'' {\em Nature},
  vol.~449, pp.~72--74, Sept. 2007.

\bibitem{Fausti2011}
D.~Fausti, R.~I. Tobey, N.~Dean, S.~Kaiser, A.~Dienst, M.~C. Hoffmann, S.~Pyon,
  T.~Takayama, H.~Takagi, and A.~Cavalleri, ``{Light-Induced Superconductivity
  in a Stripe-Ordered Cuprate},'' {\em Science}, vol.~331, pp.~189--191, Jan.
  2011.

\bibitem{mankowsky14}
R.~Mankowsky, A.~Subedi, M.~Först, S.~O. Mariager, M.~Chollet, H.~T. Lemke,
  J.~S. Robinson, J.~M. Glownia, M.~P. Minitti, A.~Frano, M.~Fechner, N.~A.
  Spaldin, T.~Loew, B.~Keimer, A.~Georges, and A.~Cavalleri, ``{Nonlinear
  lattice dynamics as a basis for enhanced superconductivity in
  YBa$_2$Cu$_3$O$_{6.5}$},'' {\em Nature}, vol.~516, pp.~71--73, Dec. 2014.

\bibitem{Miller2015}
T.~A. Miller, R.~W. Chhajlany, L.~Tagliacozzo, B.~Green, S.~Kovalev,
  D.~Prabhakaran, M.~Lewenstein, M.~Gensch, and S.~Wall, ``Terahertz field
  control of in-plane orbital order in {La$_{0.5}$Sr$_{1.5}$MnO$_4$},'' {\em
  Nature Communications}, vol.~6, p.~8175, Sept. 2015.

\bibitem{mankowsky17}
R.~Mankowsky, A.~von Hoegen, M.~F\"orst, and A.~Cavalleri, ``{Ultrafast
  Reversal of the Ferroelectric Polarization},'' {\em Phys. Rev. Lett.},
  vol.~118, p.~197601, May 2017.

\bibitem{fechner2023}
M.~Fechner, M.~Först, G.~Orenstein, V.~Krapivin, A.~S. Disa, M.~Buzzi, A.~von
  Hoegen, G.~de~la Pena, Q.~L. Nguyen, R.~Mankowsky, M.~Sander, H.~Lemke,
  Y.~Deng, M.~Trigo, and A.~Cavalleri, ``Quenched lattice fluctuations in
  optically driven {SrTiO$_3$},'' 2023.

\bibitem{gonze1997dynamical}
X.~Gonze and C.~Lee, ``Dynamical matrices, {Born} effective charges, dielectric
  permittivity tensors, and interatomic force constants from density-functional
  perturbation theory,'' {\em Phys. Rev. B}, vol.~55, pp.~10355--10368, Apr.
  1997.

\bibitem{Khalsa2021}
G.~Khalsa, N.~A. Benedek, and J.~Moses, ``{Ultrafast Control of Material
  Optical Properties via the Infrared Resonant Raman Effect},'' {\em Phys. Rev.
  X}, vol.~11, p.~021067, June 2021.

\bibitem{Caruso2023}
F.~Caruso and M.~Zacharias, ``Quantum theory of light-driven coherent lattice
  dynamics,'' {\em Phys. Rev. B}, vol.~107, p.~054102, Feb. 2023.

\bibitem{Subedi2014}
A.~Subedi, A.~Cavalleri, and A.~Georges, ``Theory of nonlinear phononics for
  coherent light control of solids,'' {\em Phys. Rev. B}, vol.~89, p.~220301,
  June 2014.

\bibitem{claassen2017}
M.~Claassen, H.-C. Jiang, B.~Moritz, and T.~P. Devereaux, ``Dynamical
  time-reversal symmetry breaking and photo-induced chiral spin liquids in
  frustrated {Mott} insulators,'' {\em Nature Communications}, vol.~8, p.~1192,
  Oct. 2017.

\bibitem{Oka2019}
T.~Oka and S.~Kitamura, ``{Floquet Engineering of Quantum Materials},'' {\em
  Annual Review of Condensed Matter Physics}, vol.~10, no.~1, pp.~387--408,
  2019.

\bibitem{delatorre21}
A.~de~la Torre, D.~M. Kennes, M.~Claassen, S.~Gerber, J.~W. McIver, and M.~A.
  Sentef, ``{Colloquium: Nonthermal pathways to ultrafast control in quantum
  materials},'' {\em Rev. Mod. Phys.}, vol.~93, p.~041002, Oct. 2021.

\bibitem{magnus2004hill}
W.~Magnus and S.~Winkler, {\em Hill's Equation}.
\newblock Dover Books on Mathematics Series, Dover Publications, 2004.

\bibitem{Rand2018}
I.~Kovacic, R.~Rand, and S.~Mohamed~Sah, ``{Mathieu's Equation and Its
  Generalizations: Overview of Stability Charts and Their Features},'' {\em
  Applied Mechanics Reviews}, vol.~70, Feb. 2018.
\newblock 020802.

\bibitem{VASP1}
G.~Kresse and J.~Hafner, ``\emph{Ab initio} molecular dynamics for liquid
  metals,'' {\em Phys. Rev. B}, vol.~47, pp.~558--561, Jan. 1993.

\bibitem{VASP2}
G.~Kresse and J.~Furthmüller, ``Efficiency of ab-initio total energy
  calculations for metals and semiconductors using a plane-wave basis set,''
  {\em Computational Materials Science}, vol.~6, pp.~15--50, July 1996.

\bibitem{VASP3}
G.~Kresse and J.~Furthm\"uller, ``Efficient iterative schemes for \emph{ab
  initio} total-energy calculations using a plane-wave basis set,'' {\em Phys.
  Rev. B}, vol.~54, pp.~11169--11186, Oct. 1996.

\bibitem{VASP_PAW}
G.~Kresse and D.~Joubert, ``From ultrasoft pseudopotentials to the projector
  augmented-wave method,'' {\em Phys. Rev. B}, vol.~59, pp.~1758--1775, Jan.
  1999.

\bibitem{Baroni2001DFPT}
S.~Baroni, S.~de~Gironcoli, A.~Dal~Corso, and P.~Giannozzi, ``Phonons and
  related crystal properties from density-functional perturbation theory,''
  {\em Rev. Mod. Phys.}, vol.~73, pp.~515--562, July 2001.

\bibitem{Vousden1951}
P.~Vousden, ``{A Study of the Unit-cell Dimensions and Symmetry of certain
  Ferroelectric Compounds of Niobium and Tantalum at Room Temperature},'' {\em
  Acta Crystallographica}, vol.~4, pp.~373--376, July 1951.

\bibitem{ISHIHARA2004}
T.~Ishihara, N.~S. Baik, N.~Ono, H.~Nishiguchi, and Y.~Takita, ``Effects of
  crystal structure on photolysis of {H$_2$O} on {K–Ta} mixed oxide,'' {\em
  Journal of Photochemistry and Photobiology A: Chemistry}, vol.~167,
  pp.~149--157, Oct. 2004.

\bibitem{Schmidbauer2012}
M.~Schmidbauer, A.~Kwasniewski, and J.~Schwarzkopf, ``{High-precision absolute
  lattice parameter determination of {SrTiO${\sb 3}$, DyScO${\sb 3}$ and
  NdGaO${\sb 3}$} single crystals},'' {\em Acta Crystallographica Section B},
  vol.~68, pp.~8--14, Feb. 2012.

\bibitem{King-Smith1993}
R.~D. King-Smith and D.~Vanderbilt, ``Theory of polarization of crystalline
  solids,'' {\em Phys. Rev. B}, vol.~47, pp.~1651--1654, Jan. 1993.

\bibitem{Resta1993}
R.~Resta, ``{Macroscopic Electric Polarization as a Geometric Quantum Phase},''
  {\em Europhysics Letters}, vol.~22, p.~133, Apr. 1993.

\bibitem{Vanderbilt1993}
D.~Vanderbilt and R.~D. King-Smith, ``Electric polarization as a bulk quantity
  and its relation to surface charge,'' {\em Phys. Rev. B}, vol.~48,
  pp.~4442--4455, Aug. 1993.

\bibitem{QuantumEspresso2009}
P.~Giannozzi, S.~Baroni, N.~Bonini, M.~Calandra, R.~Car, C.~Cavazzoni,
  D.~Ceresoli, G.~L. Chiarotti, M.~Cococcioni, I.~Dabo, A.~D. Corso,
  S.~de~Gironcoli, S.~Fabris, G.~Fratesi, R.~Gebauer, U.~Gerstmann,
  C.~Gougoussis, A.~Kokalj, M.~Lazzeri, L.~Martin-Samos, N.~Marzari, F.~Mauri,
  R.~Mazzarello, S.~Paolini, A.~Pasquarello, L.~Paulatto, C.~Sbraccia,
  S.~Scandolo, G.~Sclauzero, A.~P. Seitsonen, A.~Smogunov, P.~Umari, and R.~M.
  Wentzcovitch, ``{QUANTUM ESPRESSO}: a modular and open-source software
  project for quantum simulations of materials,'' {\em Journal of Physics:
  Condensed Matter}, vol.~21, p.~395502, Sept. 2009.

\bibitem{GARRITY2014}
K.~F. Garrity, J.~W. Bennett, K.~M. Rabe, and D.~Vanderbilt, ``Pseudopotentials
  for high-throughput {DFT} calculations,'' {\em Computational Materials
  Science}, vol.~81, pp.~446--452, Jan. 2014.

\bibitem{ISOTROPYGeneral}
H.~T. Stokes, D.~M. Hatch, and B.~J. Campbell, ``{ISOTROPY} {Software}
  {Suite}.''
\newblock iso.byu.edu.

\bibitem{ISODISTORT}
B.~J. Campbell, H.~T. Stokes, D.~E. Tanner, and D.~M. Hatch, ``{{\it
  ISODISPLACE}: a web-based tool for exploring structural distortions},'' {\em
  Journal of Applied Crystallography}, vol.~39, pp.~607--614, Aug 2006.

\bibitem{numpy}
C.~R. Harris, K.~J. Millman, S.~J. van~der Walt, R.~Gommers, P.~Virtanen,
  D.~Cournapeau, E.~Wieser, J.~Taylor, S.~Berg, N.~J. Smith, R.~Kern, M.~Picus,
  S.~Hoyer, M.~H. van Kerkwijk, M.~Brett, A.~Haldane, J.~F. del R{\'i}o,
  M.~Wiebe, P.~Peterson, P.~G{\'e}rard-Marchant, K.~Sheppard, T.~Reddy,
  W.~Weckesser, H.~Abbasi, C.~Gohlke, and T.~E. Oliphant, ``Array programming
  with {NumPy},'' {\em Nature}, vol.~585, pp.~357--362, Sept. 2020.

\bibitem{Lufaso2001}
M.~W. Lufaso and P.~M. Woodward, ``{Prediction of the crystal structures of
  perovskites using the software program {\it SPuDS}},'' {\em Acta
  Crystallographica Section B}, vol.~57, pp.~725--738, Dec 2001.

\bibitem{Sell2008}
A.~Sell, A.~Leitenstorfer, and R.~Huber, ``Phase-locked generation and
  field-resolved detection of widely tunable terahertz pulses with amplitudes
  exceeding 100 {MV/cm},'' {\em Opt. Lett.}, vol.~33, pp.~2767--2769, Dec.
  2008.

\bibitem{Knorr2017}
M.~Knorr, J.~Raab, M.~Tauer, P.~Merkl, D.~Peller, E.~Wittmann, E.~Riedle,
  C.~Lange, and R.~Huber, ``Phase-locked multi-terahertz electric fields
  exceeding 13 {MV/cm} at a 190 {kHz} repetition rate,'' {\em Opt. Lett.},
  vol.~42, pp.~4367--4370, Nov. 2017.

\bibitem{jose_1998}
J.~V. José and E.~J. Saletan, {\em Classical Dynamics: A Contemporary
  Approach}, p.~382–491.
\newblock Cambridge University Press, 1998.

\bibitem{Cartella2018}
A.~Cartella, T.~F. Nova, M.~Fechner, R.~Merlin, and A.~Cavalleri, ``Parametric
  amplification of optical phonons,'' {\em Proceedings of the National Academy
  of Sciences}, vol.~115, pp.~12148--12151, Nov. 2018.

\bibitem{forst11}
M.~F{\"o}rst, C.~Manzoni, S.~Kaiser, Y.~Tomioka, Y.~Tokura, R.~Merlin, and
  A.~Cavalleri, ``Nonlinear phononics as an ultrafast route to lattice
  control,'' {\em Nature Physics}, vol.~7, pp.~854--856, Nov. 2011.

\bibitem{forst15}
M.~F{\"o}rst, R.~Mankowsky, and A.~Cavalleri, ``{Mode-Selective Control of the
  Crystal Lattice},'' {\em Accounts of Chemical Research}, vol.~48,
  pp.~380--387, Feb. 2015.

\bibitem{khalsa18}
G.~Khalsa and N.~A. Benedek, ``Ultrafast optically induced
  ferromagnetic/anti-ferromagnetic phase transition in {GdTiO$_3$} from first
  principles,'' {\em npj Quantum Materials}, vol.~3, p.~15, Mar. 2018.

\bibitem{Disa2021}
A.~S. Disa, T.~F. Nova, and A.~Cavalleri, ``Engineering crystal structures with
  light,'' {\em Nature Physics}, vol.~17, pp.~1087--1092, Oct. 2021.

\bibitem{note4}
We note that the condensation of $Q_{\vec{q}}$ can create other Raman-active
  phonons via additional anharmonic lattice energy terms which may be accessed
  in some crystals through the conventional nonlinear phononics effect
  \cite{juraschek17}. For simplicity, we ignore these pathways in our analysis
  since, though their inclusion is straightforward, it can quickly become
  unnecessarily cumbersome for our purposes.

\bibitem{Nova19}
T.~F. Nova, A.~S. Disa, M.~Fechner, and A.~Cavalleri, ``Metastable
  ferroelectricity in optically strained {SrTiO$_3$},'' {\em Science},
  vol.~364, pp.~1075--1079, June 2019.

\bibitem{Li2019}
X.~Li, T.~Qiu, J.~Zhang, E.~Baldini, J.~Lu, A.~M. Rappe, and K.~A. Nelson,
  ``Terahertz field-induced ferroelectricity in quantum paraelectric
  {SrTiO$_3$},'' {\em Science}, vol.~364, pp.~1079--1082, June 2019.

\bibitem{Fleury1968}
P.~A. Fleury, J.~F. Scott, and J.~M. Worlock, ``{Soft Phonon Modes and the
  110$^\circ$K Phase Transition in SrTiO$_3$},'' {\em Phys. Rev. Lett.},
  vol.~21, pp.~16--19, July 1968.

\bibitem{Stephenson1908}
A.~Stephenson, ``{XX. On Induced Stability},'' {\em The London, Edinburgh, and
  Dublin Philosophical Magazine and Journal of Science}, vol.~15, pp.~233--236,
  Apr. 1908.

\bibitem{Kapitsa_1951}
P.~L. Kapitsa, ``Pendulum with vibrating suspension,'' {\em Usp. Phys.
  Sciences}, vol.~44, pp.~7--20, May 1951.

\bibitem{Mertelj2019}
T.~Mertelj and V.~V. Kabanov, ``{Comment on ``Ultrafast Reversal of the
  Ferroelectric Polarization''},'' {\em Phys. Rev. Lett.}, vol.~123, p.~129701,
  Sept. 2019.

\bibitem{Abalmasov2020}
V.~A. Abalmasov, ``Ultrafast reversal of the ferroelectric polarization by a
  midinfrared pulse,'' {\em Phys. Rev. B}, vol.~101, p.~014102, Jan. 2020.

\bibitem{Chen2022}
P.~Chen, C.~Paillard, H.~J. Zhao, J.~{\'I}{\~{n}}iguez, and L.~Bellaiche,
  ``Deterministic control of ferroelectric polarization by ultrafast laser
  pulses,'' {\em Nature Communications}, vol.~13, p.~2566, May 2022.

\bibitem{Glazer1972Notation1}
A.~M. Glazer, ``{The Classification of Tilted Octahedra in Perovskites},'' {\em
  Acta Crystallographica Section B}, vol.~28, pp.~3384--3392, Nov. 1972.

\bibitem{Glazer1975Notation2}
A.~M. Glazer, ``{Simple Ways of Determining Perovskite Structures},'' {\em Acta
  Crystallographica Section A}, vol.~31, pp.~756--762, Nov. 1975.

\bibitem{Slowly_Varying_Envelop_Approx_1965}
F.~Arecchi and R.~Bonifacio, ``{Theory of Optical Maser Amplifiers},'' {\em
  IEEE Journal of Quantum Electronics}, vol.~1, no.~4, pp.~169--178, 1965.

\bibitem{juraschek17}
D.~M. Juraschek, M.~Fechner, and N.~A. Spaldin, ``Ultrafast {Structure}
  {Switching} through {Nonlinear} {Phononics},'' {\em Phys. Rev. Lett.},
  vol.~118, p.~054101, Jan. 2017.

\end{thebibliography}
\bibliographystyle{ieeetr}


\newcommand{\beginsupplement}{%
        \setcounter{table}{0}
        \renewcommand{\thetable}{S\arabic{table}}%
        \setcounter{figure}{0}
        \renewcommand{\thefigure}{S\arabic{figure}}%
        \setcounter{section}{0}
        \renewcommand{\thesection}{S\arabic{section}}%
        \setcounter{equation}{0}
        \renewcommand{\theequation}{S\arabic{equation}}%
     }

\newpage
\onecolumngrid
\begin{center}
\textbf{Supplementary Information: Coherent control of the translational and point group symmetries of crystals with light}

Guru Khalsa$^1$, Jeffrey Z. Kaaret,$^2$ and Nicole A. Benedek$^1$

1. Department of Materials Science and Engineering, Cornell University, Ithaca NY 14853, USA

2. School of Applied and Engineering Physics, Cornell University, Ithaca NY 14853, USA

\end{center}

\beginsupplement
\section{Approximate solutions to the Floquet phase boundaries}
\label{sec:phaseboundary}

\begin{figure}[ht]
    \centering
    \includegraphics[width=.6\textwidth]{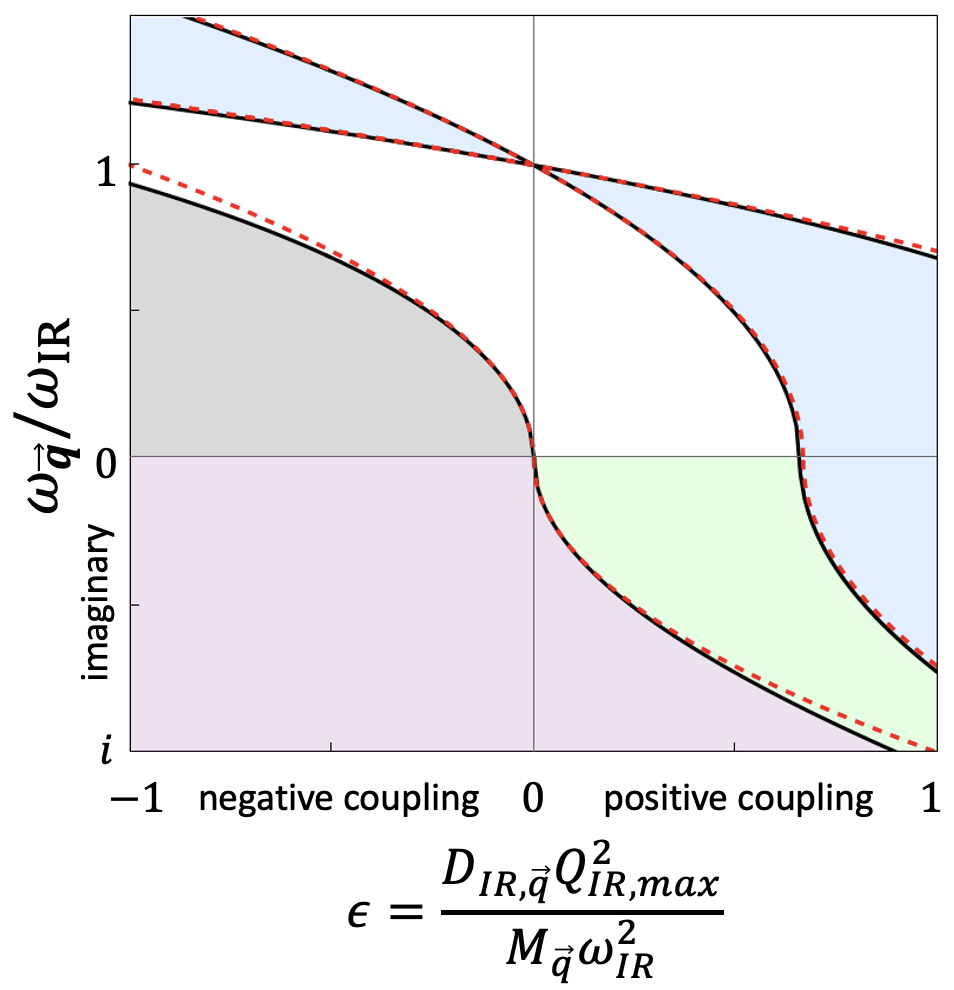}
    \caption{Floquet phase diagram with approximate analytic results for the phase boundaries (red, dashed). The analytic expressions can be found in the main text below Eqn. 7 ($|\epsilon| = \delta$), in Sec. III A 5 ($\epsilon = |\delta|$), and in Appendix C.}
    \label{fig:PhaseBoundary_approx}
\end{figure}

Approximate analytic results for the phase boundaries are shown in Fig. \ref{fig:PhaseBoundary_approx}. This suggests that the analytic results given in the main text are good approximations for the transition from trivial motion (white regions -- Region I) to novel dynamical response discussed in the main text (shaded regions), compared to those found with Floquet theory.

\begin{figure}[ht]
    \centering
    \includegraphics[width=.6\textwidth]{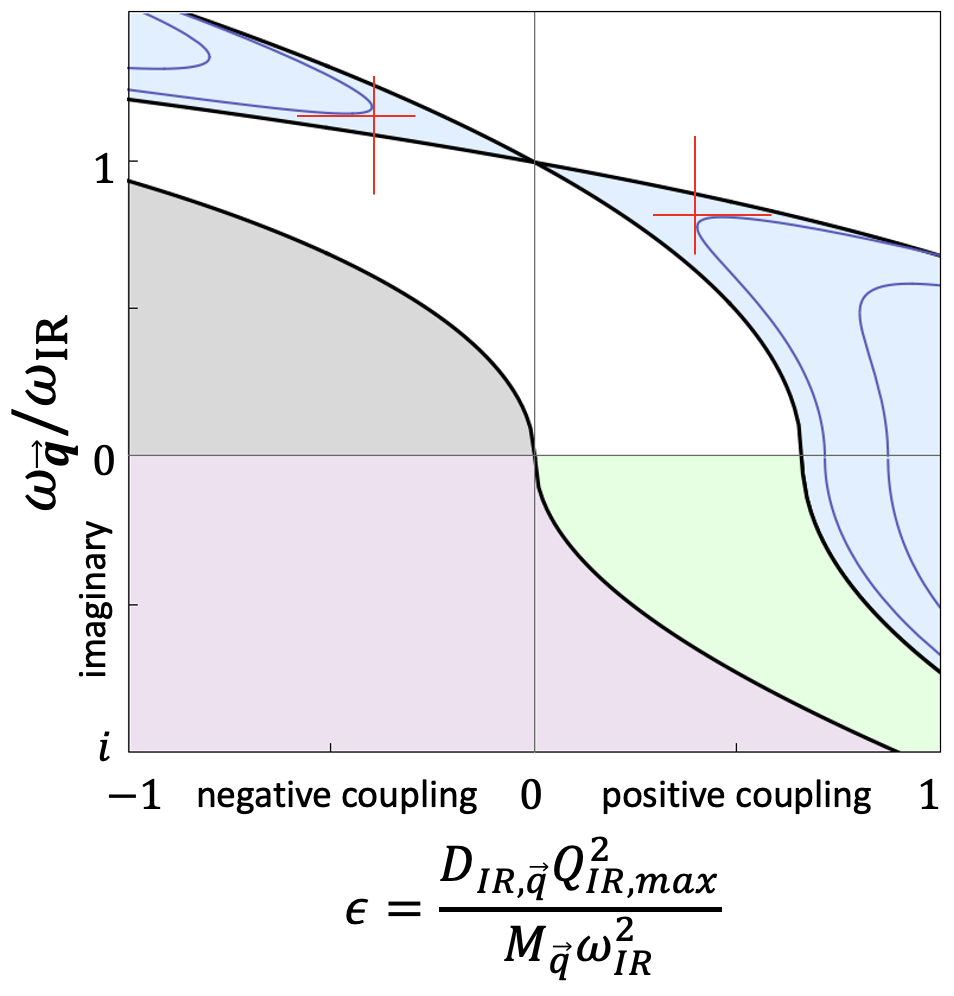}
    \caption{Floquet phase diagram with analytic phase boundaries including damping ($\nu = 0.2, 0.4$). The red lines in the figure show the threshold values of $\sqrt\delta = \omega_{\vec{q}}/\omega_{IR}$ and $\epsilon = D_{IR,\vec{q}}/M_{\vec{q}}\omega_{IR}^2$ needed to enter the parametric oscillation regime near $\sqrt{\delta} = 1$. The analytic expressions can be found in Appendix C of the main text.}
    \label{fig:Damping}
\end{figure}

Fig. \ref{fig:Damping} shows the affect of damping ($\nu=2\gamma_{\vec{q}}/\omega_{IR}$) on the phase boundary between trivial damped motion of $Q_{\vec{q}}$ (white region -- Region I) and exponentially growing parametric oscillation of $Q_{\vec{q}}$ (blue region -- Region III). The red lines in the figure show the threshold values the driving strength ($|\epsilon| \geq 2\nu $) and the frequency ratio $\sqrt\delta$ ($|1-\delta_\nu|\geq\sqrt 3 \nu$, where $\delta_\nu = \delta -\nu^2/4$) that must be overcome to see the parametric oscillatory behavior (Appendix C in the main text for further discussion).

\section{The effect of strain and linear-quadratic coupling}
\label{sec:Strain}

Large strain states are expected to be observed on timescales comparable to the phonon-phonon scattering component of the lifetime [S1]. That is, when the energy transfers from $Q_{IR}$ and $Q_{\vec{q}}$ to the acoustic branch. We note that a nearly instantaneous strain response (on the scale of the electronic scattering rate $\sim$ 1 ps) is also expected as the electrons accommodate the increased energy density from the optical field, but we expect this to be a small effect. Here, we discuss how this strain may impact the dynamics of the phase boundary, focusing on Region II. 

Including strain components in Eqn. 1 gives

\begin{equation} \label{eqn:Potential_energy_strain}
    \begin{split}
        U  &= \frac{1}{2}K_{IR}Q_{IR}^2 + \frac{1}{2}K_{\vec{q}}Q_{\vec{q}}^2 + \frac{1}{2}C_{\varepsilon}\varepsilon^2 \\
        &+ \alpha_{IR} \varepsilon Q_{IR}^2 + \alpha_{\vec{q}} \hspace{.05cm} \varepsilon Q_{\vec{q}}^2  +D_{IR,\vec{q}}Q_{IR}^2Q_{\vec{q}}^2 -\Delta \vec{P}\cdot \vec{E}.
    \end{split}
\end{equation}

\noindent
where $\varepsilon$ is a strain coordinate, $C_\varepsilon$ is an elastic stiffness, and $\alpha_{IR/\vec{q}}$ is the coupling between strain and $Q_{IR/\vec{q}}$. In this equation, $\varepsilon$ represents the strain induced by the excitation and will generally be a linear combination of the equilibrium strain tensor components. $C_\varepsilon$, and $\alpha_{IR/\vec{q}}$ are therefore effective elastic parameters associated with the induced strain response. Note that in the calculation of the strain coefficients, only the lattice constants are allowed to change. That is, the only atomic motion included is due to the IR-active phonon.

Taking the same steps as done in Sec. II A we define an effective force constant for $Q_{\vec{q}}$:
\begin{equation} \label{eqn:ZE_Force_constant_strain}
    \begin{split}
        \tilde{K}_{\vec{q}}(Q_{IR},\varepsilon)  &= K_{\vec{q}}+ 2 \alpha_{\vec{q}} \hspace{.05cm} \varepsilon   +2D_{IR,\vec{q}}Q_{IR}^2,
    \end{split}
\end{equation}

\noindent 
We find the peak strain amplitude $\varepsilon^*$ by solving $-\frac{\partial U}{\partial \varepsilon}=0$, while keeping $Q_{\vec{q}} =0$. Resulting in: ${\varepsilon^* = \alpha_{IR}  Q_{IR}^2/C_\varepsilon }$, which is substituted into Eqn. \ref{eqn:ZE_Force_constant_strain} to find,
\begin{equation} \label{eqn:ZE_Force_constant_strain_temp1}
        \tilde{K}_{\vec{q}}(Q_{IR})  = K_{\vec{q}}+ 2 \left(  \frac{\alpha_{IR} \alpha_{\vec{q}}}{C_\varepsilon}  +D_{IR,\vec{q}} \right) Q_{IR}^2.
\end{equation}
\noindent 
This highlights the qualitative finding that if $\frac{\alpha_{IR} \alpha_{\vec{q}}}{C_\varepsilon}$ has the same sign as $D_{IR,\vec{q}}$, then strain and the biquadratic coupling work in concert to decrease the critical field. If they differ in sign, strain and the biquadratic coupling work in opposition to increase the threshold field. We are unaware of any \emph{a priori} strategy for anticipating the relative sign of the coupling and therefore expect theoretical modeling will generally be necessary for this level of detail in the definition of the phase boundary. We note that this framework for including the effect of other modes in the phase boundary applies to \emph{any} linear-quadratic third-order coupling term. Therefore, the conventional nonlinear phononics effect, with anharmonic coupling $\propto Q_{IR}^2Q_{R}$ can also be treated on equal footing with strain, having only a quantitative effect on the phase boundary.

As an example of this, we focus on the Region II phase boundary for KTaO$_3$ for excitation of the 16.7 THz IR-active phonon polarized along the [100] direction. The phase boundary in the absence of strain is found to occur at 4.6 MV/cm for a 500 fs duration Gaussian pulse. We find that $\alpha_{IR} \alpha_{\vec{q}}/C_\varepsilon$ is negative. That is, strain and the biquadratic coupling work in concert to condense $Q_{\vec{q}}$. We find the phase boundary including strain to be \text{3.8 MV/cm}.

\begin{figure}[ht]
    \centering
    \includegraphics[width=.6\textwidth]{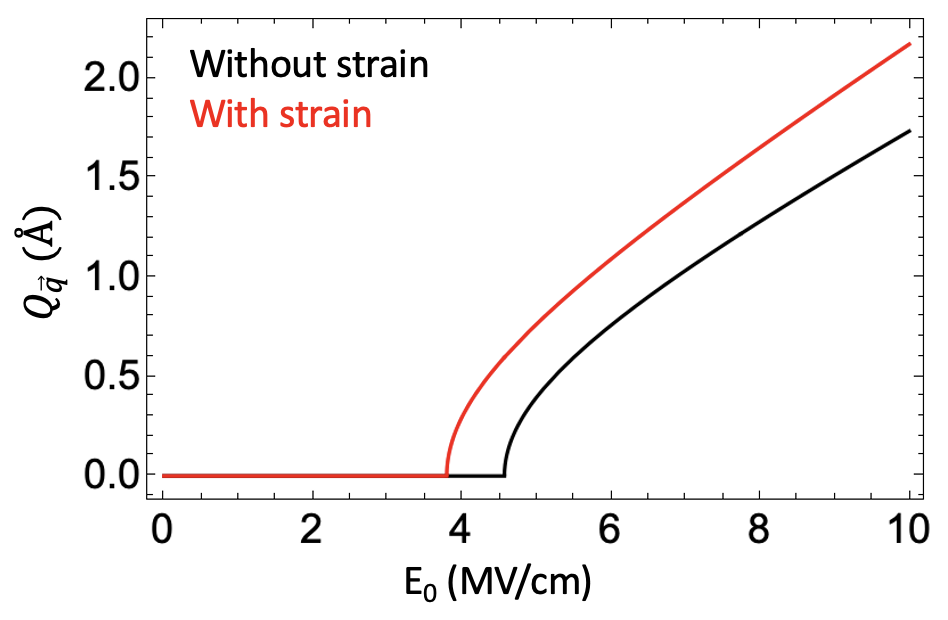}
    \caption{Condensed value of $Q_{\vec{q}}$ as a function of peak electric field for without (black) and with (red) strain for the 16.7 THz IR active phonon in KTaO$_3$ polarized along the [100] direction.}
    \label{fig:crit_strain}
\end{figure}

\section{Eigenmodes of the induced instability $Q_{\vec{q}}$}
\label{sec:eigenmodes}

The induced structural instabilities due to IR-active phonon pumping are primarily octahedral tilts/rotations, as discussed in the main text, but involve smaller distortions that are also allowed by symmetry. An eigenmode decomposition of the $Q_{\vec{q}}$'s found in Table III of the main text, referenced to the parent Pm$\bar{3}$m structure, is given in Tables \ref{supp:tab:100}, \ref{supp:tab:110}, and \ref{supp:tab:111}. The symmetry labels and directions were generated by the ISOTROPY Software Suite [S2, S3].

\begin{center}
\begin{table}[h] 
\caption{Real-space eigendisplacements of the induced instabilities for IR phonon excitation along the crystallographic [100] direction in KTaO$_3$, found in Table III of the main text. The IR phonon lowers the symmetry of the equilibrium Pm$\bar{3}$m ($O_h$) structure to P4mm ($C_{4v}$). The irreducible representation (Irrep) of $Q_{\vec{q}}$ in P4mm is A$_5$ (second column) which has two high symmetry directions shown in the top and bottom of the table. The decomposition of $Q_{\vec{q}}$ in the Pm$\bar{3}$m reference is given in the last columns. The sum of the squares of the eigenmode components is unity.}
\label{supp:tab:100}
{\setlength{\tabcolsep}{1em}
\begin{tabular}{rccrrr}
\hline \hline
\multicolumn{1}{c}{\multirow{3}{*}{\begin{tabular}[c]{@{}c@{}}Frequency\\ {[}THz{]}\end{tabular}}} & \multicolumn{1}{c}{\multirow{3}{*}{\begin{tabular}[c]{@{}c@{}}Irrep of $Q_{\vec{q}}$ in\\ space group\\ induced by $Q_{IR}$\end{tabular}}} & \multicolumn{4}{c}{} \\
 &  & \multicolumn{4}{c}{Decomposition of $Q_{\vec{q}}$ in Pm$\bar{3}$m} \\ 
 & &  &  &  &  \\ \hline 
&  & \multicolumn{1}{r}{$R_5^+[10\bar{1}]$:K}  & \multicolumn{1}{r}{$R_4^-[10\bar{1}]$:Ta} & \multicolumn{1}{r}{$R_4^+[101]$:O} & \multicolumn{1}{r}{$R_5^+[10\bar{1}]$:O} \rule{0pt}{9pt}\\

\multicolumn{1}{r}{16.7} & \multicolumn{1}{c}{\multirow{3}{*}{A$_5$(a,0)}} & \multicolumn{1}{c}{-} & \multicolumn{1}{r}{0.060726} & \multicolumn{1}{r}{0.978766} & 0.195523 \\
\multicolumn{1}{r}{5.5} &  & \multicolumn{1}{r}{-0.318172} & \multicolumn{1}{r}{-0.059278} & \multicolumn{1}{r}{0.945980} & 0.019344 \\
\multicolumn{1}{r}{3.3} &  & \multicolumn{1}{c}{-} & \multicolumn{1}{r}{0.019962} & \multicolumn{1}{r}{0.983722} & 0.178586 \\ 
 &  &  &  &  &  \\ 
 &  & \multicolumn{1}{r}{$R_5^+[100]$:K} & \multicolumn{1}{r}{$R_4^-[001]$:Ta} & \multicolumn{1}{r}{$R_4^+[100]$:O} &  \multicolumn{1}{r}{$R_5^+[100]$:O} \\ 
\multicolumn{1}{r}{16.7} & \multicolumn{1}{c}{\multirow{3}{*}{A$_5$(a,a)}} & \multicolumn{1}{c}{-} & \multicolumn{1}{r}{-0.060726} & \multicolumn{1}{r}{0.978766} & 0.195523 \\
\multicolumn{1}{r}{5.5} &  & \multicolumn{1}{r}{-0.318172} & \multicolumn{1}{r}{0.059278} & \multicolumn{1}{r}{0.945980} & 0.019344 \\
\multicolumn{1}{r}{3.3} &  & \multicolumn{1}{c}{-} & \multicolumn{1}{r}{-0.019962} & \multicolumn{1}{r}{0.983722} & 0.178586 \\ \hline \hline
\end{tabular}
}
\end{table}
\end{center}

\begin{center}

\begin{table}[h]
 \caption{Real-space eigendisplacements of the induced instabilities for IR phonon excitation along the crystallographic [110] direction in KTaO$_3$, found in Table III of the main text. The IR phonon lowers the symmetry of the equilibrium Pm$\bar{3}$m ($O_h$) structure to Amm2 ($C_{2v}$). The irreducible representation (Irrep) of $Q_{\vec{q}}$ in Amm2 is T$_4$ for the 16.7 THz and 3.3 THz phonons, and Y$_4$ for the 5.5 THz IR phonon (second column) shown in the top and bottom of the table, respectively. The decomposition of $Q_{\vec{q}}$ in the Pm$\bar{3}$m reference is given in the last columns. The sum of the squares of the eigenmode components is unity.}
\label{supp:tab:110}
{\setlength{\tabcolsep}{.7em}
\begin{tabular}{lcrrrrl}
\hline \hline
\multicolumn{1}{c}{\multirow{3}{*}{\begin{tabular}[c]{@{}c@{}}Frequency\\ {[}THz{]}\end{tabular}}} & \multicolumn{1}{c}{\multirow{3}{*}{\begin{tabular}[c]{@{}c@{}}Irrep of $Q_{\vec{q}}$ in\\ space group\\ induced by $Q_{IR}$\end{tabular}}} & \multicolumn{3}{c}{} &  &  \\
 &  & \multicolumn{5}{c}{Decomposition of $Q_{\vec{q}}$ in Pm$\bar{3}$m}  \\ 
 &  &  &  &  &  &  \\ \hline
 &  & & \multicolumn{1}{r}{$R_4^-[01\bar{1}]$:Ta} & \multicolumn{1}{r}{$R_3^+[010]$:O} & $R_4^+[100]$:O &    \\ 
\multicolumn{1}{r}{16.7} & \multicolumn{1}{c}{\multirow{2}{*}{T$_4$}} & & \multicolumn{1}{r}{-0.061179} & \multicolumn{1}{r}{-0.026906} & -0.997764 &    \\
\multicolumn{1}{r}{3.3} &  & & \multicolumn{1}{r}{-0.058514} & \multicolumn{1}{r}{-0.041206} & -0.997436 &   \\ 
 &  &  &  &  &  &  \\
 &  & \multicolumn{1}{r}{$M_5^-$:K} & \multicolumn{1}{r}{$M_5^-$:Ta} & \multicolumn{1}{r}{$M_2^+$:O} & \multicolumn{1}{r}{$M_3^+$:O} & \multicolumn{1}{r}{$M_5^-$:O} \\
 &  & \multicolumn{1}{r}{$[1\bar{1}0000]$} & \multicolumn{1}{r}{$[1\bar{1}0000]$} & \multicolumn{1}{r}{$[100]$} & \multicolumn{1}{r}{$[100]$} & \multicolumn{1}{r}{$[1\bar{1}0000]$} \\ 
\multicolumn{1}{r}{5.5} & \multicolumn{1}{c}{Y$_4$} & \multicolumn{1}{r}{0.543105} & \multicolumn{1}{r}{0.126706} & \multicolumn{1}{r}{0.026610} & \multicolumn{1}{r}{-0.816469} & \multicolumn{1}{r}{0.147150} \\ \hline \hline
\end{tabular}
}
\end{table}
\end{center}

\begin{center}
\begin{table}[h]
 \caption{Real-space eigendisplacements of the induced instabilities for IR phonon excitation along the crystallographic [111] direction in KTaO$_3$, found in Table III of the main text. The IR phonon lowers the symmetry of the equilibrium Pm$\bar{3}$m ($O_h$) structure to R3m ($C_{3v}$). The irreducible representation (Irrep) of $Q_{\vec{q}}$ in R3m is T$_2$. The decomposition of $Q_{\vec{q}}$ in the Pm$\bar{3}$m reference is given in the last column where only the $R_4^+[111]$:O, representing a pure out-of-phase octahedral tilt, is seen. The sum of the squares of the eigenmode components is unity.}
\label{supp:tab:111}

{\setlength{\tabcolsep}{1em}
\begin{tabular}{ccc}
\hline \hline
\multirow{3}{*}{\begin{tabular}[c]{@{}c@{}}Frequency\\ {[}THz{]}\end{tabular}} & \multirow{3}{*}{\begin{tabular}[c]{@{}c@{}}Irrep of $Q_{\vec{q}}$ in\\ space group\\ induced by $Q_{IR}$\end{tabular}} &  \\
 &  & \multicolumn{1}{c}{Decomposition of $Q_{\vec{q}}$ in Pm$\bar{3}$m} \\ 
 &  &  \\ \hline
  &  & $R_4^+[111]$:O \rule{0pt}{9pt} \\ 
\multicolumn{1}{r}{16.7} & \multirow{3}{*}{T$_2$} & 1.000000 \\
\multicolumn{1}{r}{5.5} &  & 1.000000 \\
\multicolumn{1}{r}{3.3} &  & 1.000000 \\ \hline \hline
\end{tabular}}
\end{table}

\clearpage
\section{Zone-edge phonon response to IR-active phonon excitation: bond-length changes}
\label{sec:bondlength}

The strong coupling between the zone-edge phonon $Q_{\vec{q}}$ and the IR-active phonon $Q_{IR}$ leading to new translational and point group symmetries in KTaO$_3$ found in Sec. III B, can be understood in terms of the geometric network of bonds (see last two paragraphs of Sec. III B). The large changes in bond lengths due to the IR phonon excitation are energetically unfavorable for the crystal. Octahedral tilts associated with zone-edge phonons provide a pathway for the crystal to mitigate this situation by increasing the shortest bond length. This is shown in Fig. \ref{fig:Bondlength} where the shortest time-averaged Ta-O bond length is shown as a function of IR-active phonon displacement (blue), with the inclusion of the R$_4^+$ octahedral tilt (red) and the full $A_5$ phonon (green). Once the threshold value of $Q_{IR}$ is reached, both octahedral tilts and the other subtle structural distortions contributing to the $A_5$ mode, increase the shortest time-averaged bond length, lowering the configuration energy of the crystal lattice.

\begin{figure}[ht]
    \centering
    \includegraphics[width=.6\textwidth]{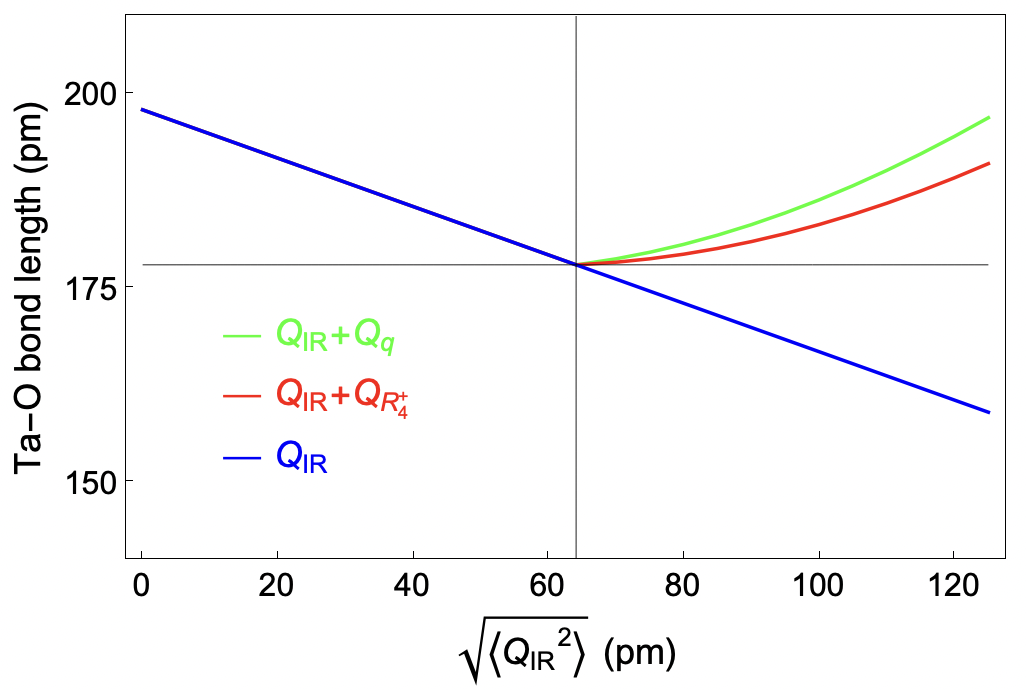}
    \caption{Change is the shortest time-averaged Ta-O bond length due to the IR phonon (blue), the IR phonon with the $R_4^+$ octahedral tilts (red), and the IR phonon with the total $Q_{\vec{q}}$ (green). The solid black lines show the threshold value of $Q_{IR}$ and the Ta-O bond length. The structural changes were found by adding the $Q_{\vec{q}}$ amplitudes predicted by Eqn. 10 with parameters from Table I into the equilibrium KTaO$_3$ crystal structure to calculate the bond lengths.}
    \label{fig:Bondlength}
\end{figure}

\end{center}

\section*{Supplementary references}
\begin{enumerate}
    \item[{[S1]}] J. R. Hortensius, D. Afanasiev, A. Sasani, E. Bousquet, and A. D. Caviglia, “Ultrafast strain engineering and coherent structural dynamics from resonantly driven optical phonons in LaAlO3,” npj Quantum Materials, vol. 5, p. 95, Dec. 2020.
    \item[{[S2]}] H. T. Stokes, D. M. Hatch, and B. J. Campbell, “ISOTROPY Software Suite.” iso.byu.edu.
    \item[{[S3]}] B. J. Campbell, H. T. Stokes, D. E. Tanner, and D. M. Hatch, “ISODISPLACE: a web-based tool for exploring structural distortions,” Journal of Applied Crystallography, vol. 39, pp. 607–614, Aug 2006.
\end{enumerate}

\end{document}